\documentclass[pra,aps,amsmath,amssymb,amsfonts,twocolumn,nofootinbib,longbibliography]{revtex4-1}
\usepackage{}
\usepackage{amssymb}
\usepackage{bm,mathrsfs}
\usepackage{graphicx}
\usepackage{epsfig}
\usepackage{amsmath,bbm}
\usepackage{amsfonts,amssymb}
\usepackage{times}
\usepackage{verbatim}
\usepackage[sort&compress]{natbib}
\usepackage{amsmath}
\usepackage{bm}
\usepackage{float}
\allowdisplaybreaks[4]
\usepackage[colorlinks,linkcolor=blue,anchorcolor=blue,citecolor=blue,urlcolor=magenta]{hyperref}
\usepackage{color}

\begin{document}
\title{Non-Markovian dynamics with a driven three-level giant atom in a semi-infinite photonic waveguide}
\author{S. J. Sun,$^{1}$ Z. Y. Li,$^{1}$ C. Cui,$^{1}$ Shuang Xu,$^{2}$ and H. Z. Shen$^{1,}$}
\email{Corresponding author: shenhz458@nenu.edu.cn\\https://orcid.org/0000-0002-4017-7367}

\affiliation{$^1$Center for Quantum Sciences and School of Physics, Northeast Normal University, Changchun 130024, China\\
$^2$College of Sciences, Northeastern University, Shenyang 110819, China}
\date{\today}

\begin{abstract}
The non-Markovian effects of open quantum systems subjected to external environments are deemed to be valuable resources in quantum optics and quantum information processing. In this work, we investigate the non-Markovian dynamics of a three-level giant atom coupling with a semi-infinite photonic waveguide through multiple coupling points and driven by a classical driving field. We derive the analytical expressions for the probability amplitudes of the driven three-level giant atom and obtain two independent conditions. We find two different types of bound states (including the static bound states and the periodic equal-amplitude oscillating bound states) and discuss the physical origins of the bound states formation. Moreover, we discuss the case of the driven three-level giant atom interacting with the infinite photonic waveguide, where there is only one purely imaginary solution (i.e., only one bound state condition exists) for its complex frequency (coming from the absence of mirror at one end of the waveguide) compared to that of a driven three-level giant atom coupling with a semi-infinite photonic waveguide. With this, we also find two different types of bound states, including the static bound state and the periodic equal-amplitude oscillating bound states. Finally, the above results are generalized to a more general model involving a semi-infinite photonic waveguide coupling with an arbitrary number of noninteracting three-level giant atoms driven by the driving fields. The proposed protocol could provide a pathway to precisely elucidate the non-Markovian dynamics of driven, multi-level giant atoms coupled to semi-infinite or infinite photonic waveguides.
\end{abstract}



\maketitle
\section{Introduction}
Quantum physics primarily focuses on investigating and harnessing light-matter interactions at the quantum scale \cite{Kockum192019}. These interactions can be effectively studied in platforms like cavity quantum electrodynamics (QED) systems \cite{Kimble1998, Chen023841, Raimond565, Chen012325, Walther1325, Liu023724}, circuit-QED systems \cite{Blais025005, Kang013233, Blais062320, Zhao042217, Wallraff431162}, and waveguide QED systems \cite{Roy0210012017,Gu12017,Sheremet210306824}. In waveguide QED systems, bandwidth constraints are mitigated by the waveguide's inherent support for continuous optical modes. These studies are usually carried out based on point atoms, where the wavelength of point-like atoms is usually much smaller than that of light \cite{Goy19031983,Leibfried2812003,Miller5512005,Haroche10832013}. Thus, when dealing with these systems we usually use the dipole approximation \cite{Walls2008} to simplify the interaction between atoms and photons. Several interesting effects have been found, such as strong photon-photon quantum correlations \cite{Shen2007, Liao2010,Xu2014,Zheng2011,Shi2011,DRoy2011,Gu023720}, radiations from a quantum emitter exponentially localized in the vicinity of the quantum emitter \cite{Lombardo2014,Burillo2017}, and controllable bi-photon bound states \cite{Wang2022,Yang2024,Costa052126}. In contrast, a hot topic is the study of a single quantum emitter coupling with an infinite or semi-infinite linear photonic waveguide \cite{Zhang08841,Li063703,zhao063708,wang053720,yan013301,wang013279,Chen13678}, where the single quantum emitter is modeled as a two-level emitter \cite{Fang53053845,Zhang390323352020,Tufarelli0138202013,Tufarelli45012113} or a driven three-level emitter \cite{Barkemeyer1900078}. The semi-infinite linear photonic waveguide is a one-dimensional infinite photonic waveguide with one end behaving as a perfect mirror \cite{Song6959,Fang53053845,Zhang390323352020,Tufarelli0138202013,Tufarelli45012113,Barkemeyer1900078,Dorner023816,Xin2022,Bradford063830,Chang0243052020,Fang043035043035,Zeng0303052023,Barkemeyer0237082022,Crowder0137142022,Barkemeyer0337042021,Kabuss102016,Ding230606373,Ding230716876,Wang19002252019,Wang111772019,Sinha043603}.
For example, by thinning one end of the waveguide so that it is almost transparent and connecting the other end to an opaque medium, single-ended and quasi-one-dimensional structures can be realized \cite{Claudon1742010,Bleuse1036012011,Reimer7372012,Bulgarini1211062012}. As a result, light impinging on the end always experiences some partial backreflection. Truly, it is a process of coherent quantum feedback that leads to the information backflow from the non-Markovian environment, i.e., non-Markovianity \cite{Chalabi063832,Guimond03389,
Kofman41353,John12772,Liu052139,Fong023842,Garmon010102}. Since some amount of radiation emitted by the quantum emitter is reflected back by the mirror to the quantum emitter, a decay of the quantum emitter to its ground state is inhibited to a nonzero constant due to the emergence of atom-photon bound state \cite{Xin2022,He17994}.

With deeper research in quantum optics and huge advances in superconducting circuit technology \cite{Manenti9752017,Kockum2019,Krantz0213182019,You2011}, an interesting paradigm of quantum optics dubbed \textit{giant atom} development in recent period of time characterizes cases where artificial atoms interact with surface acoustic waves or microwave photons beyond the standard dipole approximation \cite{Kockum2021,Du0237052023,Soro0137102023,Soro023712,Cilluffo0430702020,Guo2017,Longhi30172020,Du0231982022,Taylor2020,Zhao0538552020,Wang0436022021,Xiao802022,Yin0637032022,Lizy2024,Cheng0335222022,Vadiraj2021}. With the development of modern nanotechnology, people can implement giant atoms (working in the microwave regime) that interact with the surface acoustic waves (SAW) \cite{Gustafsson2014,Andersson2019}. Due to the low velocity of SAW, the wavelength of sound is no longer considered to be large compared to the size of the giant atom for a given frequency. An alternative way to engineer giant atom coupling is by meandering the one-dimensional transmission line such that the atom interacts with the photonic waveguide at multiple points \cite{Kockum2018,Kannan2020}. Moreover, the level structure of the giant atom can be engineered to form a two-level system or a three-level system \cite{Yang17994,Sun17351,Gao12414,Lin16217,Zhou055202,Ma7852024,gao013716,Sun129561,Huang050506,Weng08834,Liu18402,You00264}. Giant atoms distinguish from conventional small atoms  in at least two points: quantum interference \cite{Zhu0437102022,Regidor023030,Pichler093601,Qiu2242122023,Johansson083603,Kockum2014} and non-Markovian feedback \cite{Guo2017,Ask0138402019,Carollo2020,Kockum2018,Vega0535222021,Ask201115077,Zhao425062022,Du2236022022,Zhang220503674,Noachtar0137022022,Du123012023,Wang0437032022,Brianso0637172022,Regidor0337192023,Luo230908856,Wang230903663,Santos0536012023,Yang1151042021,Wang0350072022,Wang75802022,Liu742023,Burillo0137092020,Leonforte10275,Zhang230316480,Li35982023,Chen0431352023,Du0450102023,Kuo230707949,Jia230402072,Yin230314746,Liu1068542023,Zheng043030,Ingelsten10879,Guo0337112024,Tudela2036032019}. In the past, a lot of research on non-Markovian systems was based on point-like atoms in the traditional quantum optics framework, where the existence of mirror or multiple point-like atoms is usually required \cite{Eschner4952001,Guimond0440122017,Calajo0736012019,Milonni10961974,Zheng1136012013,Ballestero0730152013,Laakso1836012014,Ballestero0423282014,Guimond0238082016,Ramos0621042016,Dinc2132019}. For the non-Markovian giant atom, more novel phenomena can appear, e.g., persistent oscillating bound states \cite{Johansson2020,Lim0237162023,Johansson2020Xu} and non-exponential spontaneous emissions \cite{Andersson2019,Carollo2020,Roccati063603}.

However, for the driven three-level giant atom coupled with a semi-infinite (or infinite) photonic waveguide, studying the formation of bound states and deriving analytical expressions for atomic probability amplitudes have not been investigated. Therefore, these considerations motivate us to explore the non-Markovian dynamics caused by the bound states for a driven three-level giant atom coupling with a semi-infinite (or infinite) photonic waveguide.

In this work, we study the non-Markovian dynamics of a three-level giant atom coupling with a one-dimensional photonic waveguide through multiple coupling points and driven by a classical driving field. We start from the model of a driven three-level giant atom that interacts with the one-dimensional semi-infinite photonic waveguide. We derive the analytical solutions for the atomic probability amplitudes, which show nonexponential dissipations due to the transfer of the photon between multiple coupling points and its re-absorption after being reflected by the mirror. We discuss two independent conditions and physical origins for the formation of bound states. According to the number of bound states in the semi-infinite photonic waveguide, the bound states are classified into two groups: (i) the static case with one bound state; (ii) the periodic equal-amplitude oscillation with two bound states. We emphasize the effects of time delay and the driving field on bound state formation. Next, we discuss the case of infinite photonic waveguide. Due to the absence of mirror at one end of the photonic waveguide, there is only one purely imaginary solution for the complex frequency leading to this case compared to that of a driven three-level giant atom coupling with a semi-infinite photonic waveguide, which implies that there is only one bound state condition. Under this condition, we also find two different kinds of bound states, including the static bound state and the periodic equal-amplitude oscillating bound states. Finally, we further generalize the model to a more general quantum system containing many noninteracting three-level giant atoms interacting with a semi-infinite photonic waveguide via multiple coupling points.

The present paper is organized as follows. In Sec.~\ref{Sec2}, we introduce the model Hamiltonian of three-level giant atom with multiple coupling points and derive Schr\"{o}dinger equation for the operators of giant atom and bosonic field in the semi-infinite photonic waveguide. In Sec.~\ref{Sec3}, we derive a set of delay differential equations for the evolution of the driven three-level giant atom. In Sec.~\ref{Sec4}, we calculate two independent conditions for the formation of bound states. In Sec.~\ref{Sec5}, we obtain two different types of bound states and give the dynamical expressions of bound states. In Sec.~\ref{Sec7}, we discuss the case of the driven three-level giant atom coupling with a one-dimensional infinite photonic waveguide. In Sec.~\ref{Sec8}, we generalize the above results to driven noninteracting three-level giant atoms coupling with a semi-infinite photonic waveguide. In Sec.~\ref{Sec9}, we finally draw our conclusions.
\begin{figure}[t]
\centerline{
\includegraphics[width=8.9cm, height=3.6cm, clip]{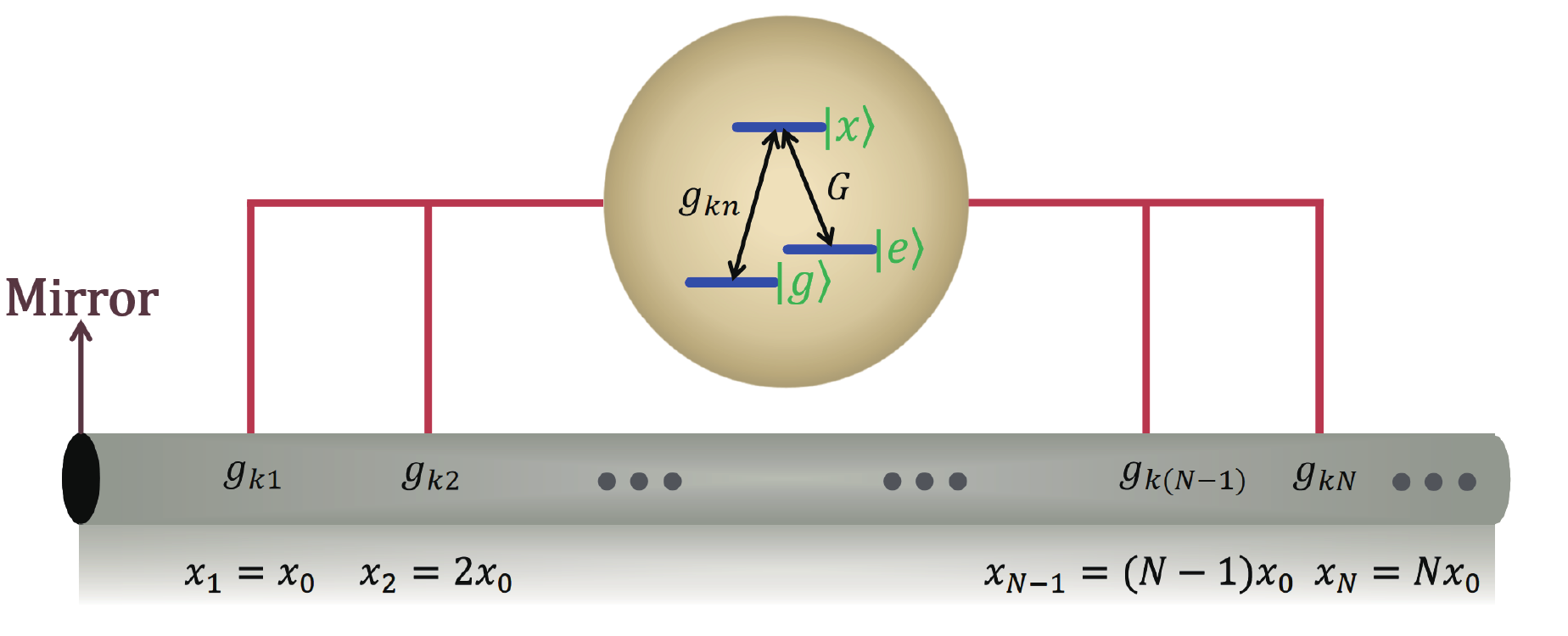}}
\vspace{-0.1cm}
\caption{(Color online) Schematic of the setup. A $\Lambda$-type three-level giant atom (driven by a classical driving field) couples with a one-dimensional semi-infinite photonic waveguide (terminated by a perfect mirror) through $N$ coupling points. The transition between the ground state $|g\rangle$ (the eigenfrequency $\omega_g$ is set to be zero) and the excited state $|x\rangle$ (the eigenfrequency $\omega_x$) is coupled to a photonic waveguide positioned at $x_n$ with the coupling coefficient $g_{kn}$ ($n = 1, 2,\cdot \cdot \cdot,N$). The distance between the adjacent coupling points is $x_0$, which leads to $x_n=nx_0$. The transition between the excited state $|x\rangle$ and the metastable state $|e\rangle$ (the eigenfrequency $\omega_e$) is driven by a laser field with the frequency $\omega_l$ and the driving strength $G$.}\label{model}
\end{figure}

\section{THE MODEL AND METHODS}\label{Sec2}

As shown in Fig.~\ref{model}, we consider a single driven $\Lambda$-type three-level giant atom coupling with a one-dimensional semi-infinite photonic waveguide along the $x$ axis via $N$ coupling points \cite{Lodahl654, Wang420183, Wang16870}. The configuration of a giant transmon coupling with a transmission line has been demonstrated in Ref.\cite{Gu12017,Peropadre0638342011,Mok0538612020,Boch1705012014,Eichler0321062012}. The ground, metastable, and excited states are denoted by levels $|g\rangle$, $|e\rangle$, and $|x\rangle$.

For the semi-infinite photonic waveguide terminated by a perfect mirror (reflectivity $R = 1$) at one end of the photonic waveguide $x=0$, we introduce the feedback from the mirror \cite{Fang53053845, Tufarelli0138202013, Zhang390323352020, Tufarelli45012113, Barkemeyer1900078}, where the part of emitted photon of the atom will transfer between the mirror and each coupling point through the photonic waveguide. The photon dispersion relation for the semi-infinite photonic waveguide with wave vector $k$ is linear \cite{Roy0210012017,Shen3020012005,Shen0238372009,Shen0238382009,Shen2130012005}, where $\upsilon$ is the photon group velocity.
The annihilation (creation) operator is denoted by ${\hat a_k}$ $(\hat a_k^\dag )$, which satisfies $[{\hat a_k},\hat a_{k'}^\dag] = \delta(k - k')$. The infinite photonic waveguide corresponds to two orthogonal standing modes with spatial profiles $ \propto \cos (kx)$ and $ \propto \sin (kx)$, respectively.

Since there is a perfect mirror at one end of the photonic waveguide, only the sine-like modes are considered. Therefore, the three-level giant atom interacts with the photonic waveguide at the $n$th coupling point with strength ${g_{kn}} \propto \sin (kn{x_0})$ ($n=1, 2 ,\cdot \cdot \cdot,N$).
Under the rotating-wave approximation (RWA), we write the total system Hamiltonian as ($\hbar\equiv1$ throughout this work)
\begin{equation}\begin{aligned}
\hat{H}_{1}=&\begin{aligned}\omega_x|x\rangle\langle x|+\omega_e|e\rangle\langle e|+\int_0^{k_c}dk\Omega_k\hat{a}_k^\dagger\hat{a}_k\end{aligned}
\\
&+Ge^{i\omega_lt}\hat\sigma_{ex}+Ge^{-i\omega_lt}\hat\sigma_{xe} \\
&+\sum_{n=1}^{N}\int_0^{k_c}dkg_{kn}(\hat{a}_k^\dagger\hat\sigma_{gx}+\hat{a}_k\hat\sigma_{xg}),\label{hatH1}
\end{aligned}\end{equation}
where ${k_c}$ represents a cutoff wave vector depending on the specific photonic waveguide. $ \hat\sigma _{df} = \left| d \right\rangle \left\langle f \right|$ $(d,f = x,g,e)$ is the raising or lowering operator of the atom, where $\omega_x$ and $\omega_e$ are eigenfrequencies corresponding to levels $|x\rangle$ and $|e\rangle$, respectively. In this case, we have set the ground state energy $\omega_g$ to be zero, i.e., $\omega_g=0$. The first line of Eq.~(\ref{hatH1}) represents the free Hamiltonian for the atom and photonic waveguide of the total system. The second line of Eq.~(\ref{hatH1}) corresponds to the interaction Hamiltonian between the three-level giant atom and external driving field with the driving strength $G$ and the frequency $\omega_l$. The last line of Eq.~(\ref{hatH1}) models the interaction Hamiltonian between the giant atom and semi-infinite photonic waveguide, where $g_{kn}$ denotes the coupling strength between the $n$th coupling point and the $k$th mode in the semi-infinite photonic waveguide. Making the unitary transformation ${\hat U} = {e^{\frac{{ - i{\omega _l}t}}{2}({{\hat \sigma }_{xx}} - {{\hat \sigma }_{ee}}+\int_0^{{k_c}} {\hat a_k^\dag } {\hat a_k}dk)}}$, the total Hamiltonian (\ref{hatH1}) of the system can be written as
\begin{equation}\begin{aligned}
\hat{H}_{2}=&\mu_1 \hat\sigma_{xx}+\mu_2\hat\sigma_{ee}
\\
&+\int_{-\infty}^{+\infty}(\Omega_{k}-\frac{\omega_{l}}{2})\hat{a}_k^\dagger\hat{a}_kdk+G(\hat\sigma_{ex}+\hat\sigma_{xe})\\ &+\sum_{n=1}^{N}\int_{-\infty}^{+\infty}dkg_{kn}(\hat{a}_k^\dagger\hat\sigma_{gx}+\hat{a}_k\hat\sigma_{xg}),
\label{hatH2}\end{aligned}\end{equation}
where $\mu_1=\omega_{x}-\frac{\omega_{l}}{2}$ and $\mu_2=\omega_{e}+\frac{\omega_{l}}{2}$. As we work under the RWA, the relevant processes involve only around $k=k_0$ and hence we can set the lower limit as negative infinity, i.e., $\int_0^{k_c}dk\to\int_{-\infty}^{+\infty}dk$ \cite{Zhang390323352020,Roy0210012017,Shen3020012005,Liao0630042016}.
Since the RWA guarantees that there is only one excitation either in the atom or in the photonic waveguide when the total system initially has one excitation, we study the non-Markovian dynamics of the giant-atom in the single-excitation subspace of the total system. Considering the total system initially being prepared in the state $\left| {{\psi (0)}} \right\rangle  = C_x (0)\left| {x,0} \right\rangle  + C_e (0)\left| {e,0} \right\rangle $, the state of the total system at time $t$ can be described by
\begin{equation}\begin{aligned}
\left| {{\psi (t)}} \right\rangle  = C_x (t)\left| {x,0} \right\rangle  + C_e (t)\left| {e,0} \right\rangle  + \int {dk\beta _k (t)} \left| {g,{1_k}} \right\rangle,
\label{psit}\end{aligned}\end{equation}
where ${C_x}(t)$ and ${C_e}(t)$ stand for the atomic probability amplitudes in the excited state $\left| {x} \right\rangle$ and the metastable state $\left|  {e} \right\rangle$, respectively. ${\beta _k}(t)$ denotes the $k$th field amplitude with ${\beta _k}(0)=0$. $\left| 0 \right\rangle $ and $\left| {{1_k}} \right\rangle$ respectively describe the vacuum state and single-photon state in the field mode with frequency ${\Omega _k}$, where the integral term in Eq.~(\ref{psit}) represents the state of a single photon propagating in the photonic waveguide. These probability amplitudes in Eq.~(\ref{psit}) can be determined by Schr\"{o}dinger equation $\frac{\partial}{\partial t}\left| {{\psi (t)}} \right\rangle=-i\hat H_2\left| {{\psi(t)}} \right\rangle$, which causes
\begin{small}
\begin{eqnarray}
\dot{C}_{x}(t)&=&-i\mu_1C_{x}(t)-iGC_{e}(t)
-i\sum_{n=1}^{N}\int_{-\infty}^{+\infty}dkg_{kn}\beta_{k}(t),
\label{dotCx}\\
\dot{C}_{e}(t)&=&-i\mu_2 C_{e}(t)-iGC_{x}(t),
\label{dotCe}\\
\dot{\beta}_{k}(t)&=&-i(\Omega_{k}-\frac{\omega_{l}}{2})\beta_{k}(t)-iC_{x}(t)\sum_{n=1}^{N}g_{kn}.
\label{dotbetak}
\end{eqnarray}
\end{small}

In the following, we will set the interaction strength $g_{kn}=\sqrt{\Gamma\upsilon/\pi}\sin(knx_0)$ \cite{Tufarelli45012113}, where $\Gamma$ represents the decay rate (in the no mirror case).

\section{Non-Markovian DYNAMICS for THE driven three-level giant atom}\label{Sec3}
The evolution time is less than the time delay corresponding to the Markovian approximation, which depends on the driving strength and the choice of parameters. When the evolution time is longer than the time delay, the photons reabsorbed by atoms show the non-Markovian memory effect. We now obtain the following equation by integrating Eq.~(\ref{dotbetak})
\begin{equation}
\begin{aligned}
\beta_{k}(t)=&-i\sqrt{\frac{\Gamma v}{\pi}}\sum_{n=1}^{N}\sin(knx_0)\\
&\cdot\int_0^{t}C_{x}(s)e^{-i(\Omega_{k}-\frac{\omega_{l}}{2})(t-s)}ds.
\label{betak}
\end{aligned}
\end{equation}
By substituting Eq.~(\ref{betak}) into Eq.~(\ref{dotCx}), we can get the set of atomic excitation probability amplitudes
\begin{align}
\dot{C}_{x}(t)=&-i\mu_1 C_{x}(t)-iGC_{e}(t)\nonumber\\
&-\frac{\Gamma}{2}\sum_{m,n=1}^{N}C_{x}(t-|m-n|\tau_0)\nonumber\\
&\cdot e^{i\frac{\omega_{l}}{2}|m-n|\tau_0}\Theta(t-\left|m-n\right|\tau_0)\nonumber\\
&+\frac{\Gamma}{2}\sum_{m,n=1}^{N}C_{x}[t-(m+n)\tau_0]\nonumber\\
&\cdot e^{i\frac{\omega_{l}}{2}(m+n)\tau_0}\Theta[t-(m+n)\tau_0],
\label{dotcx}\\
\dot{C}_{e}(t)=&-i\mu_2 C_{e}(t)-iGC_{x}(t),
\label{dotce}
\end{align}
where $\Theta(x)$ [$\Theta(x)=0$ for $x<0$ and $\Theta(x)=1$ for $x>0$] is the Heaviside step function, which describes time-delayed feedback from the reflection of the semi-infinite photonic waveguide and the coupling points. $\tau_0=x_0/v$ denotes the time for the photon to travel between any two adjacent coupling points. The first and second terms on the right-hand side of Eq.~(\ref{dotcx}) describe atomic coherent dynamics. The third term on the right-hand side of Eq.~(\ref{dotcx}) shows the transfer of photons from the $m$th coupling point to the $n$th coupling point without specular reflection, including the Markovian approximation for the re-absorption of photons released from the $m$th coupling point without mirror reflection. The last term on the right-hand side of Eq.~(\ref{dotcx}) represents the non-Markovian feedback effect where the photon released from the $m$th coupling point is absorbed by the $n$th coupling point after reflecting through the mirror. At the evolution time $t\geq|m-n|\tau_0$ and $t\geq(m+n)\tau_0$, the atomic reabsorption of the emitted photon denoted respectively in the third and last terms of Eq.~(\ref{dotcx}) occurs. The derivation details of Eqs.~(\ref{betak})-(\ref{dotce}) can be found in Appendix \ref{A}. If the driving strength $G$ and the driving frequency $\omega_l$ are zero plus $m=n=1$ (i.e., two-level point-like atom), the forms of probability amplitudes in Eqs.~(\ref{dotcx}) and (\ref{dotce}) will return back to Refs.\cite{Tufarelli0138202013,Tufarelli45012113} for the point-like atom coupled with the semi-infinite photonic waveguide.

By substituting $\hat{F}(x)$ into Eq.~(\ref{psit}), we can obtain $\hat F(x)\left|{\psi(t)}\right\rangle=\phi(x,t)\left|{g,0}\right\rangle $, where the annihilation operator of the photon in real space is $\hat F(x)=\sqrt {2/\pi}\int{dk}{{\hat a}_k}\sin(kx)$. Thus, the probability amplitude $\phi(x,t)=\sqrt{2/\pi}\int{dk}\beta_{k}(t)\sin(kx)$ of the real space field is given by
\begin{align}
\phi\left(x,t\right)=&-i\sqrt{\frac{\Gamma}{2v}}\sum_{n=1}^{N}C_{x}(t-|\tau-n\tau_0|)\nonumber\\
&\cdot e^{i\frac{\omega_{l}}{2}|\tau-n\tau_0|}\Theta(t-|\tau-n\tau_0|)\nonumber\\
&+i\sqrt{\frac{\Gamma}{2v}}\sum_{n=1}^{N}C_{x}[t-(\tau+n\tau_0)]\nonumber\\
&\cdot e^{i\frac{\omega_{l}}{2}(\tau+n\tau_0)}\Theta[t-(\tau+n\tau_0)], \label{phi}
\end{align}
where $\tau=x/v$ represents the time delay taken by a photon to perform a round trip between any two adjacent coupling points. In order to solve the atomic excitation probability amplitudes $C_{x}(t)$ and $C_{e}(t)$ determined by Eqs.~(\ref{dotcx}) and (\ref{dotce}), we apply the Laplace transformation $\tilde{C}_f(s) \equiv \int_0^{\infty} \mathrm{d} t C_f(t) e^{-s t}$ (with $f$ representing $x$ or $e$) and obtain
\begin{small}
\begin{eqnarray}
s\tilde{C}_{x}\left(s)-C_{x}(0\right) &=&-i\mu_1\tilde{C}_{x}(s)-i G\tilde{C}_{e}\left(s\right)\nonumber\\
&& -\frac{\Gamma}{2}\sum_{m,n=1}^{N}\tilde{C}_{x}(s)e_{}^{-(s-i\frac{\omega_{l}}{2})\left|m-n\right|\tau_0}\nonumber\\
&&  +\frac{\Gamma}{2}\sum_{m,n=1}^{N}\tilde{C}_{x}(s)e^{-(s-i\frac{\omega_{l}}{2})(m+n)\tau_0}
,\label{10}\\
s\tilde{C}_{e}(s)-C_{e}(0)&=&-i\mu_2 \tilde{C}_{e}(s)-iG\tilde{C}_{x}\left(s\right)
,\label{11}
\end{eqnarray}
\end{small}
which lead to
\begin{align}
\tilde{C}_{x}\left(s\right)=&\frac{C_{x}\left(0)(s+i\mu_2\right)-iGC_{e}\left(0\right)}{[s+i\mu_1+\chi_{1}(s)](s+i\mu_2)+G^2}
\label{tildeCxs},\\
\tilde{C}_{e}\left(s\right)=&\frac{C_{e}(0)[s+i\mu_1+\chi_{1}(s)]-iGC_{x}\left(0\right)}{[s+i\mu_1+\chi_{1}(s)](s+i\mu_2)+G^2}
\label{tildeCes},
\end{align}
where $\chi_{1}(s)$ is given by

\begin{align}
\chi_{1}(s)=&\frac{\Gamma}{2}\sum_{m,n=1}^{N}\left[e^{-(s-i\omega_{l}/2)|m-n|\tau_0} \right.\nonumber\\
&\left.-e^{-(s-i\omega_{l}/2)(m+n)\tau_0}\right]. \label{kai1se}
\end{align}

Obviously, the open dynamics of the system is strongly influenced by the time delay $\tau_0$, the driving strength $G$, the driving frequency $\omega_l$, the rescaling parameters $\mu_1$ and $\mu_2$, which can be found in Eqs.~(\ref{tildeCxs}) and (\ref{tildeCes}). Because of the mirror's feedback, the light that has been emitted in the past can interfere with light emitted in the present, which is sufficient for non-exponential atomic decay and non-Markovian dynamics to occur. It is worth emphasizing that no Markovian approximation is performed in deriving Eq.~(\ref{dotcx}). Thus, our scheme is in a sense more general, i.e., systems described in Eq.~(\ref{hatH1}) can be studied under Markovian and non-Markovian regimes, separately.

The time evolution of $C_{x}(t)$ and $C_{e}(t)$ can be obtained by performing the inverse Laplace transformation to Eqs.~(\ref{tildeCxs}) and~(\ref{tildeCes})
\begin{align} C_{x}(t)&=\sum_{\alpha} e^{s_{\alpha}t} \operatorname{Res}(\tilde{C}_{x}(s), s_{\alpha})\nonumber \\
&=\sum_{\alpha} e^{s_ {\alpha}t}\lim _{s\rightarrow s_{\alpha}}\tilde{C}_{x}(s)(s-s_{\alpha})\nonumber, \\
C_{e}(t)&=\sum_{\alpha} e^{s_{\alpha}t}\operatorname{Res}(\tilde{C}_{e}(s), s_ {\alpha})\nonumber \\
&=\sum_{\alpha} e^{s_{\alpha}t}\lim _{s\rightarrow s_{\alpha}}\tilde{C}_{e}(s)(s-s_{\alpha}), \end{align}
where $\operatorname{Res}(\tilde{C}_{f}(s), s_{\alpha})$ is the residue of $\tilde{C}_{f}(s)$ at the pole $ s_ {\alpha}$ with $f$=$x$ or $e$. The poles of $\tilde{C}_{x}(s)$ and $\tilde{C}_{e}(s)$ are obtained by the complex roots (marked $s_{\alpha}$) of the following equation

\begin{align}
&[s_{\alpha}+i\mu_1+\frac{\Gamma}{2}\sum_{m,n=1}^{N}e^{-(s_{\alpha}-i\omega_{l}/2)\left|m-n\right|\tau_0}\nonumber\\
&-\frac{\Gamma}{2}\sum_{m,n=1}^{N}e^{-(s_{\alpha}-i\omega_{l}/2)\left(m+n\right)\tau_0}](s_{\alpha}+i\mu_2)+G^2=0,\nonumber\\
\label{sk}
\end{align}
which gives all complex roots (including the pure imaginary roots $s_j=-i\omega_j$ with the real number $\omega_j$, see the discussions in Sec.~\ref{Sec4}). For finite time delay $\tau_0>0$, Eq.~(\ref{sk}) holds more than one solution.

Thus, we have the time-dependent solutions for Eqs.~(\ref{dotcx}) and (\ref{dotce}) by performing the inverse Laplace transform on Eqs.~(\ref{tildeCxs}) and (\ref{tildeCes}) as follows
\begin{small}\begin{align}
C_{x}(t)=&\sum_{j}\frac{[C_{x}\left(0)(s_{j}+i\mu_2\right)-iGCe(0)]e^{-i\omega_{j}t}}{[1+\chi_{2}(s_j)](s_{j}+i\mu_2)+[s_{j}+i\mu_1+\chi_{1}(s_j)]}\nonumber\\
&+\sum_{\alpha}\frac{[C_{x}\left(0)(s_{\alpha}+i\mu_2\right)-iGCe(0)]e^{s_{\alpha}t}}{[1+\chi_{2}(s_{\alpha})](s_{\alpha}+i\mu_2)+[s_{\alpha}+i\mu_1+\chi_{1}(s_{\alpha})]},
\label{Cxtsk}
\end{align}\end{small}
\begin{small}\begin{align}
C_{e}(t)=&\sum_{j}\frac{\{ C_{e}(0)[s_{j}+i\mu_1+\chi_{1}(s_j)]-iGC_{x}(0)\}e^{-i\omega_{j}t}}{[1+\chi_{2}(s_j)](s_{j}+i\mu_2)+[s_{j}+i\mu_1+\chi_{1}(s_j)]}\nonumber\\
&+\sum_{\alpha}\frac{\{ C_{e}(0)[s_{\alpha}+i\mu_1+\chi_{1}(s_{\alpha})]-iGC_{x}(0)\}e^{s_{\alpha}t}}{[1+\chi_{2}(s_{\alpha})](s_{\alpha}+i\mu_2)+[s_{\alpha}+i\mu_1+\chi_{1}(s_{\alpha})]}, \label{Cetsk}
\end{align}\end{small}
with
\begin{align}
\chi_{2}(s)=&\frac{\Gamma}{2}\sum_{m,n=1}^{N}\left[(m+n)\tau_0e^{-(s-i\omega_{l}/2)(m+n)\tau_0} \right. \nonumber\\
&\left.-|m-n|\tau_0e^{-(s-i\omega_{l}/2)|m-n|\tau_0}\right],\label{kai2se}
\end{align}
where the first term of Eqs.~(\ref{Cxtsk}) and (\ref{Cetsk}) oscillates with time (which denotes the formation of the bound states), while the second term oscillates damping (tending to zero in long-time limit) originating from the complex roots $s_{\alpha}$.

\section{DISCUSSION OF BOUND STATE CONDITIONS}\label{Sec4}
In this section, we give two independent conditions for the creation of bound states and discuss whether these conditions can coexist. Usually, the complex frequency $s_{\alpha}$ in Eq.~(\ref{sk}) has a negative real part, which denotes the relaxation rate. In some particular situations, $s_{\alpha}$ can be purely imaginary. In this case, the corresponding mode is a bound state that does not decay despite the dissipative environment. We seek the pure imaginary solution $s_j=-i\omega_j$ ($\omega_j$ denotes the real number) in Eq.~(\ref{sk}), whose dynamics corresponds to the first term of Eqs.~(\ref{Cxtsk}) and (\ref{Cetsk}). Thus, we divide Eq.~(\ref{sk}) into imaginary and real parts respectively described through
\begin{align}
&\frac{1}{2} \Gamma\csc ^2\left(\frac{\tilde{\omega}_j \tau_0}{2}\right)\left(2 \omega_e-2 \omega_j+\omega_l\right)\sin ^2\left(\frac{N\tilde{\omega}_j\tau_0}{2}\right)\nonumber \\
&\cdot\sin ^2\left[\frac{(N+1) \tilde{\omega}_j \tau_0}{2}\right]=0,\label{im}\\
&G^2-\frac{1}{16}\csc ^2\left(\frac{\tilde{\omega}_j \tau_0}{2}\right)\left(2 \omega_e-2 \omega_j+\omega_l\right)\nonumber\\
&\cdot\{\Gamma(2N+1) \sin(\tilde{\omega}_j \tau_0)-2 \Gamma \sin(N\tilde{\omega}_j \tau_0)\nonumber\\
&-2 \Gamma \sin[(N+1) \tilde{\omega}_j \tau_0]+\Gamma \sin[(2N+1) \tilde{\omega}_j \tau_0]\nonumber\\
&-4\sin ^2\left(\frac{\tilde{\omega}_j \tau_0}{2}\right)(2\omega_j-2 \omega_x+\omega_l)\}=0,\label{re}
\end{align}
where $\tilde{\omega}_{j}=(2\omega_{j}+\omega_{l})/2$. More details of the calculations in Eqs.~(\ref{im}) and (\ref{re}) can be found in Appendix \ref{B}. Solving Eqs.~(\ref{im}) and~(\ref{re}) simultaneously, we obtain two independent solutions
\begin{align}
s_{j} & =-i\omega_{j}=-i\frac{2j\pi}{N\tau_0}+i\frac{\omega_{l}}{2}\label{skN},
\end{align}
and
\begin{align}
s_{j} & =-i\omega_{j}=-i\frac{2j\pi}{(N+1)\tau_0}+i\frac{\omega_{l}}{2}\label{skNz1},
\end{align}
where the corresponding two bound state conditions respectively satisfy
\begin{align}
\omega_{x}\tau_0&=\frac{2j\pi}{N}-\frac{1}{2}N\Gamma\tau_0\cot\left(\frac{j\pi}{N}\right)+\frac{\tau_0G^2}{\omega_{e}+\omega_{l}-\frac{2j\pi}{N\tau_0}},
\label{omegak1}
\end{align}
and
\begin{align}
\omega_{x}\tau_0&=\frac{\tau_0G^2}{\omega_{e}+\omega_{l}-\frac{2j\pi}{\left(N+1\right)\tau_0}}\notag\\
&\quad +\frac{2j\pi}{N+1}-\frac{1}{2}\left(N+1\right)\Gamma\tau_0\cot\left(\frac{j\pi}{N+1}\right)
\label{omegak2},
\end{align}
equivalently
\begin{align}
\omega_{e}\tau_0&=\frac{2\tau_0G^2}{\Gamma N\cot\left(\frac{j\pi}{N}\right)-\frac{4j\pi}{N\tau_0}+2\omega_{x}}-\omega_{l}\tau_0+\frac{2j\pi}{N}, \label{bs1}
\end{align}
and
\begin{align}
\omega_{e}\tau_0&=-\omega_{l}\tau_0+\frac{2j\pi}{N+1} \notag\\
&\quad +\frac{2\tau_0G^2}{\Gamma(N+1)\cot\left(\frac{j\pi}{N+1}\right)-\frac{4j\pi}{(N+1)\tau_0}+2\omega_{x}}\label{bs2},
\end{align}
with $j$ taking the integer number.

\begin{figure}[t]
\centerline{
\includegraphics[width=8.8cm, height=6.5cm, clip]{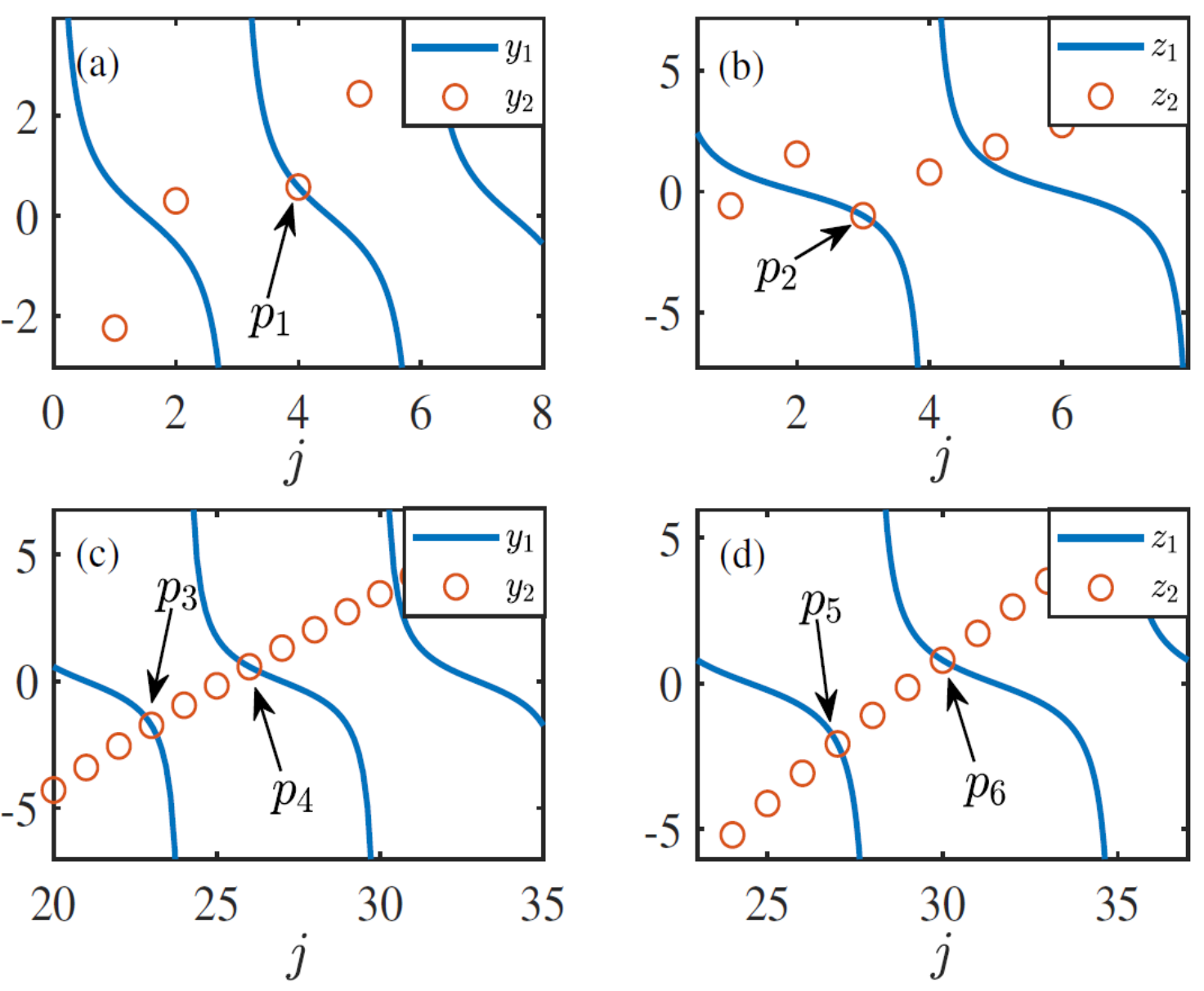}}
\vspace{-0.3cm}
\caption{The blue-solid line in Fig.~\ref{y1y2}(a) and (c) corresponds to $y_1=\cot(j\pi/N)$, while the red circle represents $y_2=4j\pi/\Gamma N^2\tau_0-2\omega_x/\Gamma N+G^2/\Gamma(N\omega_e+N\omega_l-2j\pi/\tau_0)$, where $j$ takes the integer number. The intersections between $y_1$ and $y_2$ are the solutions of the transcendental equation in Eq.~(\ref{omegak1}). The values of $j$ for the points in Fig.~\ref{y1y2}(a)-(d) are $j_{p_1}=4$, $j_{p_2}=3$, $j_{p_3}=23$, $j_{p_4}=26$, $j_{p_5}=27$, and $j_{p_6}=30$, respectively. The other parameters chosen are (a) $N=3$, $G \tau_0= 0.6\pi$, $\omega_x \tau_0=2\pi$, $\omega_e \tau_0=0.5818\pi$, $\omega_l \tau_0=1.2 \pi$, and $\Gamma \tau_0=0.3 \pi$; (c) $N=6$, $G \tau_0= 2.8\pi$, $\omega_x \tau_0=7.0366\pi$, $\omega_e \tau_0=0.2\pi$, $\omega_l \tau_0=2.5 \pi$, and $\Gamma \tau_0=0.1825\pi$. The blue-solid line in Fig.~\ref{y1y2}(b) and (d) corresponds to $z_1=\cot[j\pi/(N+1)]$, while the red circle represents $z_2=4j\pi/\Gamma(N+1)^2\tau_0-2\omega_x/\Gamma(N+1)+G^2/\Gamma[(N+1)\omega_e+(N+1)\omega_l-2j\pi/\tau_0]$. The solutions of the transcendental equation given by Eq.~(\ref{omegak2}) occur at the intersections between $z_1$ and $z_2$. The other parameters chosen are (b) $N=3$, $G \tau_0= 0.5\pi$, $\omega_x \tau_0=1.2\pi$, $\omega_e \tau_0=0.5222\pi$, $\omega_l \tau_0=0.7\pi$, and $\Gamma \tau_0=0.3 \pi$; (d) $N=6$, $G \tau_0= 2.8\pi$, $\omega_x \tau_0=6.9350\pi$, $\omega_e \tau_0=02\pi$, $\omega_l \tau_0=2.5 \pi$, and $\Gamma \tau_0=0.1079\pi$.}\label{y1y2}
\end{figure}

In the Markovian limit $\Gamma \tau_0 \rightarrow 0$, the bound-state conditions in Eqs.~(\ref{omegak1}) and~(\ref{omegak2}) are simplified as
\begin{align}
\omega_{x}\tau_0&=\frac{2j\pi}{N}+\frac{\tau_0G^2}{\omega_{e}+\omega_{l}-\frac{2j\pi}{N\tau_0}},\label{Momegak1}
\end{align}
and
\begin{align}
\omega_{x}\tau_0&=\frac{2j\pi}{N+1}+\frac{\tau_0G^2}{\omega_{e}+\omega_{l}-\frac{2j\pi}{\left(N+1\right)\tau_0}}\label{Momegak2}.
\end{align}
In the non-Markovian regime ($\Gamma \tau_0 > 0$), the additional nonlinear cotangent terms in bound-state conditions cannot be neglected.
For Eq.~(\ref{omegak1}), we define $y_1=\cot(j\pi/N)$ and $y_2=4j\pi/\Gamma N^2\tau_0-2\omega_x/\Gamma N+G^2/\Gamma(N\omega_e+N\omega_l-2j\pi/\tau_0)$. With Eq.~(\ref{omegak2}), we have $z_1=\cot[j\pi/(N+1)]$ and $z_2=4j\pi/\Gamma(N+1)^2\tau_0-2\omega_x/\Gamma(N+1)+G^2/\Gamma[(N+1)\omega_e+(N+1)\omega_l-2j\pi/\tau_0]$. We respectively plot the figure of functions $y_1$, $y_2$, $z_1$, and $z_2$ for $j$, which are shown in Fig.~\ref{y1y2}. The $j$ values of the intersections in Fig.~\ref{y1y2} are $j_{p_1}=4$, $j_{p_2}=3$, $j_{p_3}=23$, $j_{p_4}=26$, $j_{p_5}=27$, and $j_{p_6}=30$, respectively. From Fig.~\ref{y1y2}, we get that at most two bound states can coexist by changing the parameters. Two scenarios are obtained based on the number of bound states in the semi-infinite photonic waveguide and are summarized as follows (more details are shown in Sec. \ref{Sec5}).

(i) One of the bound state conditions in Eqs.~(\ref{omegak1}) and~(\ref{omegak2}) is satisfied, and there is only one solution. In this case, there are two cases given by Eqs.~(\ref{CxAnaN})-(\ref{CeAnaN}) and~(\ref{CxAnaN1})-(\ref{CeAnaN1}).

(ii) Two integers $j_1$ and $j_2$ satisfy the bound state condition for Eq.~(\ref{omegak1}) or Eq.~(\ref{omegak2}). In this case, there are two cases given by Eqs.~(\ref{2CxAnaN})-(\ref{2CeAnaN}) and~(\ref{2CxAnaN1})-(\ref{2CeAnaN1}).
\\
\\
\section{BOUND STATES IN THE SEMI-INFINITE PHOTONIC WAVEGUIDE}\label{Sec5}
In this section, we investigate the non-Markovian dynamics in the two cases obtained from the discussion of the bound state conditions in Sec.~\ref{Sec4}. We categorize them into the case of steady population containing only one bound state frequency and the case of periodic equal-amplitude oscillations where two bound states coexist. In the following subsections, we will separately discuss these two cases and give the physical origin for the formation of bound states.

\subsection{ No inversion of population}
We consider the static bound state satisfying case (i) in Sec.~\ref{Sec4}, which implies that the bound state in the semi-infinite photonic waveguide has only one frequency. Substituting $s_j$ in Eq.~(\ref{skN}) into Eqs.~(\ref{Cxtsk}) and~(\ref{Cetsk}) (more details are shown in Appendix \ref{C}), the long-time atomic excitation probability amplitudes read (corresponding to the first term of Eqs.~(\ref{Cxtsk}) and (\ref{Cetsk}) with $s_j$ satisfying Eq.~(\ref{omegak1}), while the second term of Eqs.~(\ref{Cxtsk}) and (\ref{Cetsk}) tends to zero at $t\rightarrow \infty$)
\begin{widetext}\begin{align}
C_{x}(t)&\approx\frac{[(-i\frac{2j\pi}{N\tau_0}+i\frac{\omega_{l}}{2}+i\mu_2)C_{x}(0)-iGC_{e}(0)] e^{-i(\frac{2j\pi}{N\tau_0}-\frac{\omega_{l}}{2})t}}{[1+\frac{\Gamma}{2}\frac{N\tau_0}{\sin^2(j\pi/N)}](-i\frac{2j\pi}{N\tau_0}+i\frac{\omega_{l}}{2}+i\mu_2)+[-i\frac{2j\pi}{N\tau_0}+i\frac{\omega_{l}}{2}+i\mu_1+\frac{\Gamma}{2}(\frac{2N}{1-e^{i2j\pi/N}}-N)]}
\label{CxAnaN},
\\
C_{e}(t)&\approx\frac{\{[-i\frac{2j\pi}{N\tau_0}+i\frac{\omega_{l}}{2}+i\mu_1+\frac{\Gamma}{2}(\frac{2N}{1-e^{i2j\pi/N}}-N)] C_{e}(0)-iGC_{x}(0)\} e^{-i(\frac{2j\pi}{N\tau_0}-\frac{\omega_{l}}{2})t}}{[1+\frac{\Gamma}{2}\frac{N\tau_0}{\sin^2(j\pi/N)}](-i\frac{2j\pi}{N\tau_0}+i\frac{\omega_{l}}{2}+i\mu_2)+[-i\frac{2j\pi}{N\tau_0}+i\frac{\omega_{l}}{2}+i\mu_1+\frac{\Gamma}{2}(\frac{2N}{1-e^{i2j\pi/N}}-N)]}
,\label{CeAnaN}
\end{align}
while substituting $s_j$ in Eq.~(\ref{skNz1}) into Eqs.~(\ref{Cxtsk}) and~(\ref{Cetsk}) gives (corresponding to Eq.~(\ref{omegak2}))
\begin{align}
C_{x}(t)&\approx\frac{\{[-i\frac{2j\pi}{(N+1)\tau_0}+i\frac{\omega_{l}}{2}+i\mu_2]C_{x}(0)-iGC_{e}(0)\} e^{-i[\frac{2j\pi}{(N+1)\tau_0}-\frac{\omega_{l}}{2}]t}}{\{1+\frac{\Gamma}{2}\frac{(N+1)\tau_0}{\sin^2[j\pi/(N+1)]}\}[-i\frac{2j\pi}{(N+1)\tau_0}+i\frac{\omega_{l}}{2}+i\mu_2]+\{-i\frac{2j\pi}{(N+1)\tau_0}+i\frac{\omega_{l}}{2}+i\mu_1+\frac{\Gamma}{2}[\frac{2(N+1)}{1-e^{i2j\pi/(N+1)}}-N-1]\}}
,\label{CxAnaN1}\\
C_{e}(t)&\approx\frac{\{[-i\frac{2j\pi}{(N+1)\tau_0}+i\frac{\omega_{l}}{2}+i\mu_1+\frac{\Gamma}{2}\frac{2N+2}{1-e^{i2j\pi/(N+1)}}-\frac{\Gamma}{2}N-\frac{\Gamma}{2}] C_{e}(0)-iGC_{x}(0)\} e^{-i[\frac{2j\pi}{(N+1)\tau_0}-\frac{\omega_{l}}{2}]t}}{\{1+\frac{\Gamma}{2}\frac{(N+1)\tau_0}{\sin^2[j\pi/(N+1)]}\}[-i\frac{2j\pi}{(N+1)\tau_0}+i\frac{\omega_{l}}{2}+i\mu_2]+\{-i\frac{2j\pi}{(N+1)\tau_0}+i\frac{\omega_{l}}{2}+i\mu_1+\frac{\Gamma}{2}[\frac{2(N+1)}{1-e^{i2j\pi/(N+1)}}-N-1]\}}
\label{CeAnaN1},
\end{align}\end{widetext}
where $j$ is one solution satisfying the transcendental equations in Eqs.~(\ref{omegak1}) and (\ref{omegak2}) (see Fig.~\ref{y1y2}(a) and (b)).
\begin{figure}[t]
\centerline{
\includegraphics[width=8.8cm, height=6.5cm, clip]{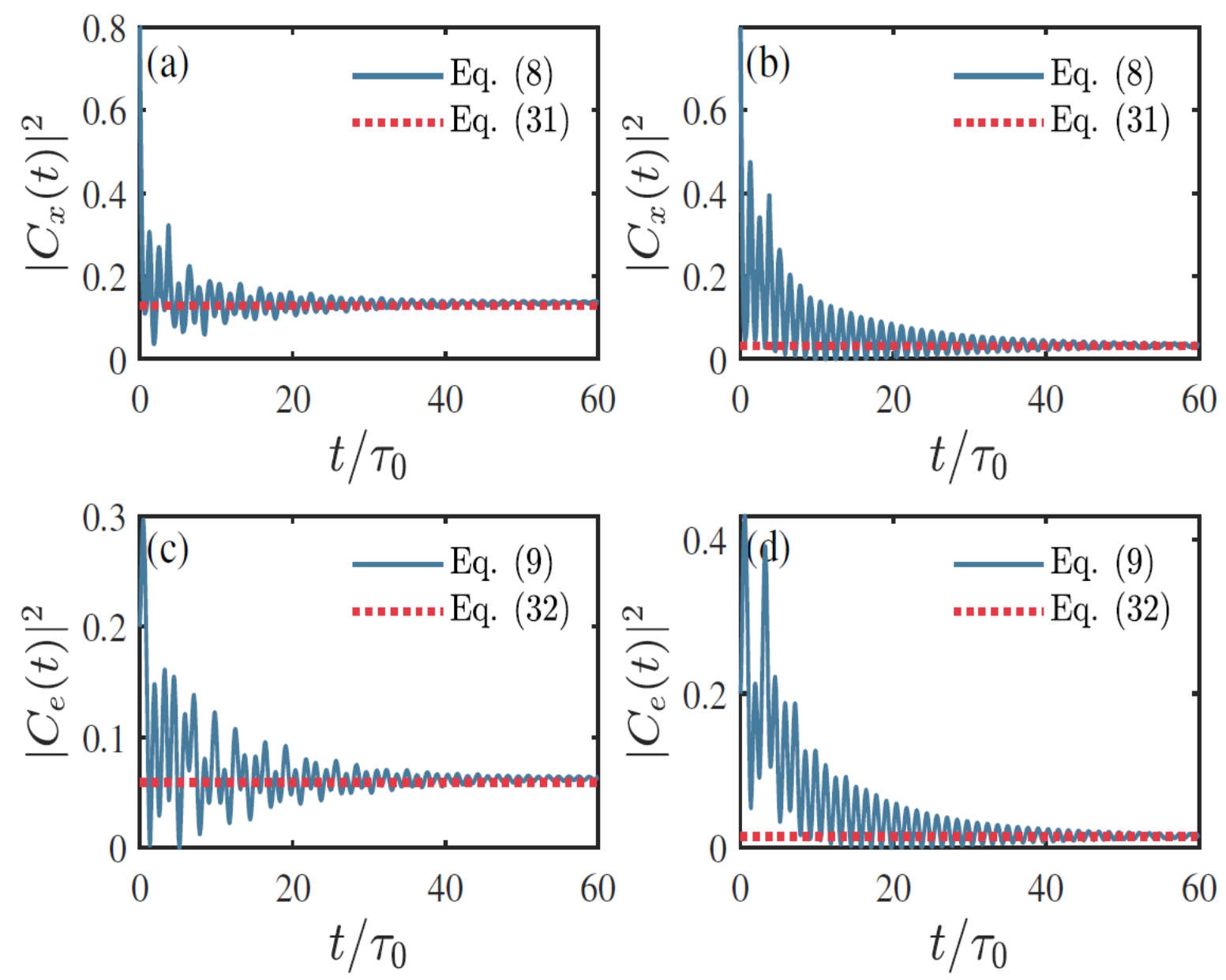}}
\vspace{-0.01cm}
\caption{(Color online) Static bound states for the three-level giant atom with the number of coupling points $N=3$. $j=4$ is chosen from Fig.~\ref{y1y2}(a). The red-dashed line corresponds to the analytical solutions in Eqs.~(\ref{CxAnaN}) and~(\ref{CeAnaN}), while the blue-solid line represents the numerical simulation with Eqs.~(\ref{dotcx}) and~(\ref{dotce}), respectively. In Fig.~\ref{StaticBS}(a) and (c), we take $G \tau_0= 0.6\pi ¦£$. In Fig.~\ref{StaticBS}(b) and (d), we take $G \tau_0= -0.6\pi ¦£$. The other parameters chosen are $\omega_x \tau_0=2\pi$, $\omega_e \tau_0=0.5818\pi$, $\omega_l \tau_0=1.2 \pi$, $\Gamma \tau_0=0.3 \pi$, ${C_x}(0) = \sqrt {0.8} $, and ${C_e}(0) = \sqrt {0.2} $.}\label{StaticBS}
\end{figure}
\begin{figure}[t]
\centerline{
\includegraphics[width=8.8cm, height=6.5cm, clip]{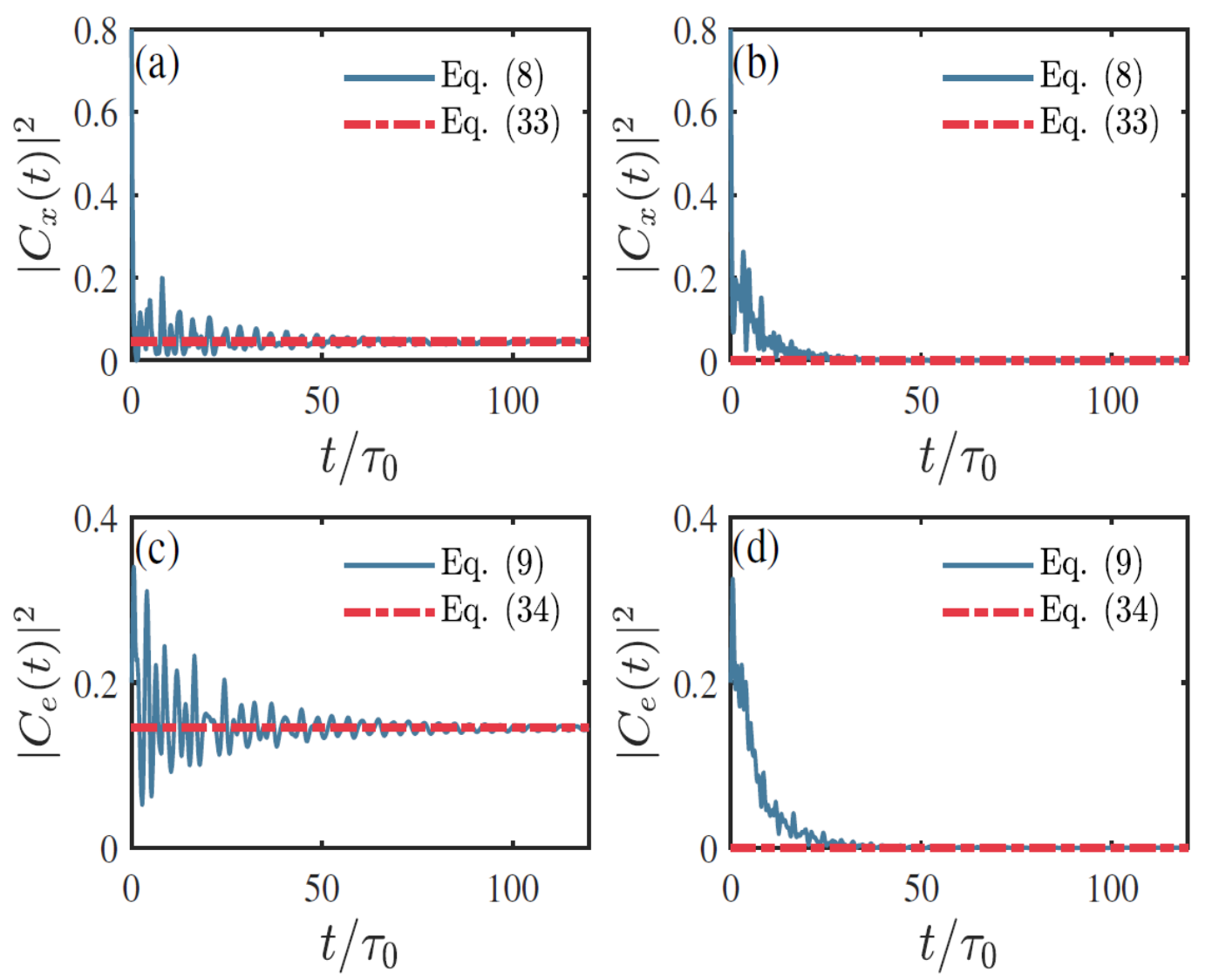}}
\vspace{-0.01cm}
\caption{(Color online) Static bound states for the three-level giant atom with the number of coupling points $N=3$. $j=3$ is chosen from Fig.~\ref{y1y2}(b). The red-dashed line corresponds to the analytical solutions in Eqs.~(\ref{CxAnaN1}) and~(\ref{CeAnaN1}), while the blue-solid line denotes the numerical simulation with Eqs.~(\ref{dotcx}) and~(\ref{dotce}), respectively. In Fig.~\ref{OneSolutionN1}(a) and (c), we take $G \tau_0= 0.5\pi ¦£$. In Fig.~\ref{OneSolutionN1}(b) and (d), we take $G\tau_0= -0.5\pi ¦£$. The other parameters chosen are $\omega_x \tau_0=1.2\pi$, $\omega_e \tau_0=0.5222\pi$, $\omega_l \tau_0=0.7 \pi$, $\Gamma \tau_0=0.3 \pi$, ${C_x}(0) = \sqrt {0.8} $, and ${C_e}(0) = \sqrt {0.2} $.}\label{OneSolutionN1}
\end{figure}
In order to compare the analytical results of Eqs.~(\ref{CxAnaN}) and~(\ref{CeAnaN}) with the numerical ones of Eqs.~(\ref{dotcx}) and~(\ref{dotce}), in Fig.~\ref{StaticBS}(a) and (b), we plot the time evolution of the giant-atom excitation probability with time $t$ for $N=3$, where $j=4$ chosen is the same as Fig.~\ref{y1y2}(a). We take $G\tau_0  = 0.6\pi $ in Fig.~\ref{StaticBS}(a) and (c), and $G\tau_0  = -0.6\pi $ in Fig.~\ref{StaticBS}(b) and (d). The blue-solid line indicates the numerical results, while the red-dashed line denotes the analytical results. It can be seen that the analytical and numerical results match well under different parameters. Similarly, in order to compare the analytical results of Eqs.~(\ref{CxAnaN1}) and~(\ref{CeAnaN1}) with the numerical ones of Eqs.~(\ref{dotcx}) and~(\ref{dotce}), we plot in Fig.~\ref{OneSolutionN1} the population dynamics for the driven three-level giant-atom ($N=3$) with respect to time $t$. The parameter $j=3$ chosen is the same as Fig.~\ref{y1y2}(b). Here, we also can find the analytical (red-dashed line) and numerical (blue-solid line) results agree well under different parameters. The probability amplitudes of driven three-level giant atom are less than those in the non-Markovian regime ($\Gamma \tau_0 > 0$). Moreover, the driven three-level giant atom excitation probabilities $|C_{x}(t)|^2$ and $|C_{e}(t)|^2$ in both cases remain at non-zero steady values after a long time, which means that the photons are caught and form bound states. The formation of the bound states arises from the transfer of the photon at multiple coupling points and the reflection from the mirrors at the endpoints of the semi-infinite photonic waveguide. Moreover, we note that the atomic probability amplitudes can be modulated by the driving strength $G$.

\begin{figure}[t]
\centerline{
\includegraphics[width=8.6cm, height=6.4cm, clip]{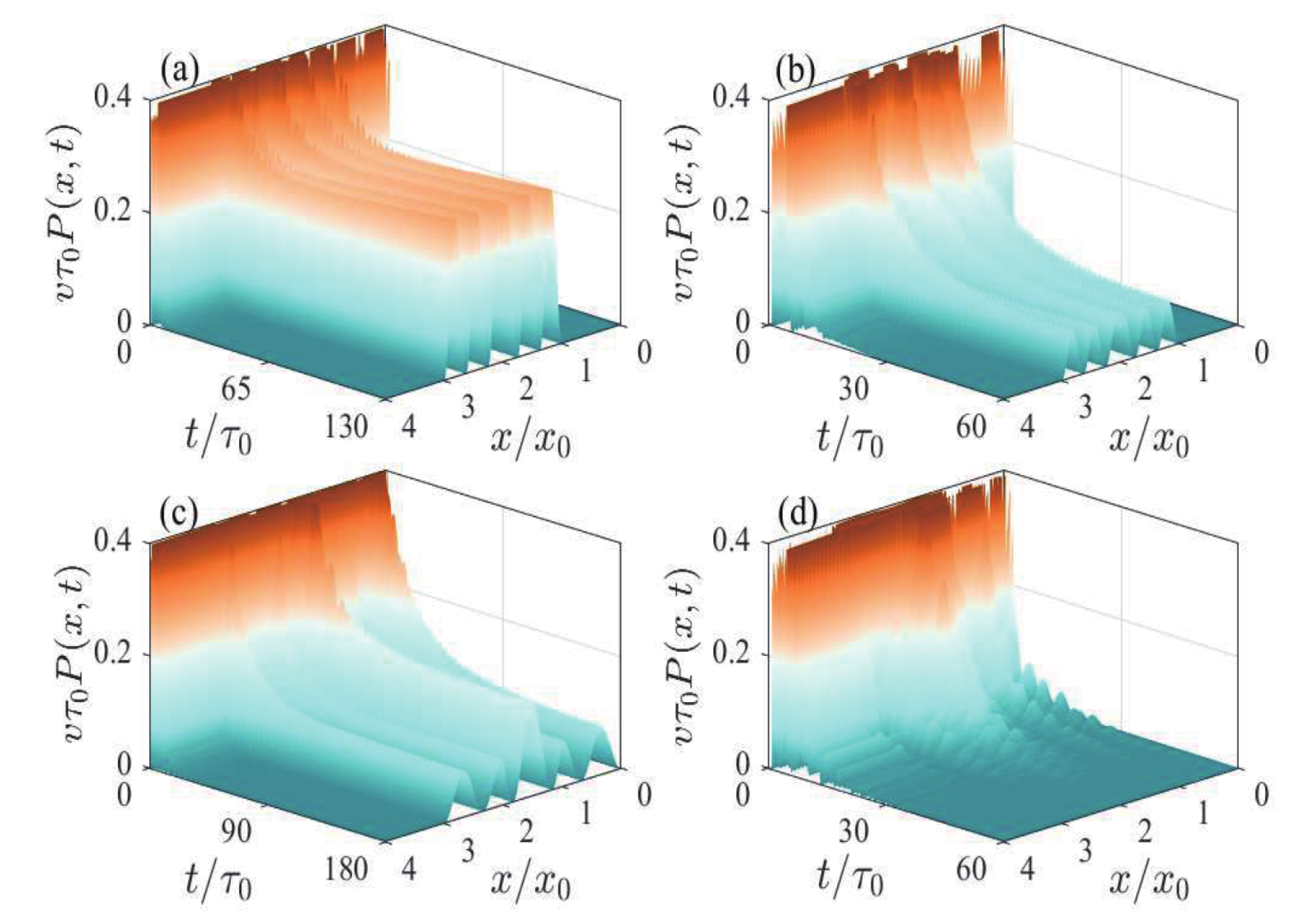}}
\vspace{-0.01cm}
\caption{(Color online) The field intensity $P(x,t) = |\phi(x,t)|^2$ in  Eq.~(\ref{phi}) as a function of time $t$ and position $x$. In Fig. \ref{3D}(a) and (b), we take $G\tau_0= 0.6\pi, -0.6\pi$, respectively, while $j=4$ is chosen from Fig.~\ref{y1y2}(a). In Fig.~\ref{3D}(c) and (d), we take $G\tau_0= 0.5\pi, -0.5\pi$, respectively, while $j=3$ is obtained by Fig.~\ref{y1y2}(b). The parameters chosen are (a) and (b) $\omega_x \tau_0=2\pi$, $\omega_e \tau_0=0.5818\pi$, and $\omega_l \tau_0=1.2 \pi$; (c) and (d) $\omega_x \tau_0=1.2\pi$, $\omega_e \tau_0=0.5222\pi$, and $\omega_l \tau_0=0.7 \pi$. The other parameters chosen are $\Gamma \tau_0=0.3 \pi$, ${C_x}(0) = \sqrt {0.8} $, and ${C_e}(0) = \sqrt {0.2} $. The top row corresponds to bound state condition in Eq.~(\ref{bs1}), while the bottom row represents bound state condition in Eq.~(\ref{bs2}).}\label{3D}
\end{figure}

In Fig.~\ref{3D}, we show how the bound state of the total system is formed. We take the field strength function $P(x,t)$ as a function of time $t$ and position $x$ and show the evolution of the field strength function with time for different parameters for both cases at $N=3$. When the position $x$ is the region $x>3x_0$, the field intensity completely vanishes. While the field intensity reaches a steady value at the region $x<3x_0$, a static bound state is formed in the photonic waveguide. This is in stark contrast to the bound state due to local impurity (small atom) in the environment, where the field bound state is exponentially suppressed on both sides of impurity (small atom). Here, we can clearly see that for the bound state condition in Eq.~(\ref{omegak2}) when the parameters are chosen as in Fig.~\ref{3D}(d), there is no bound state. This suggests that the formation of the bound state can be manipulated by mediating the driving field intensity $G$.

\begin{figure}[t]
\centerline{
\includegraphics[width=8.8cm, height=6.5cm, clip]{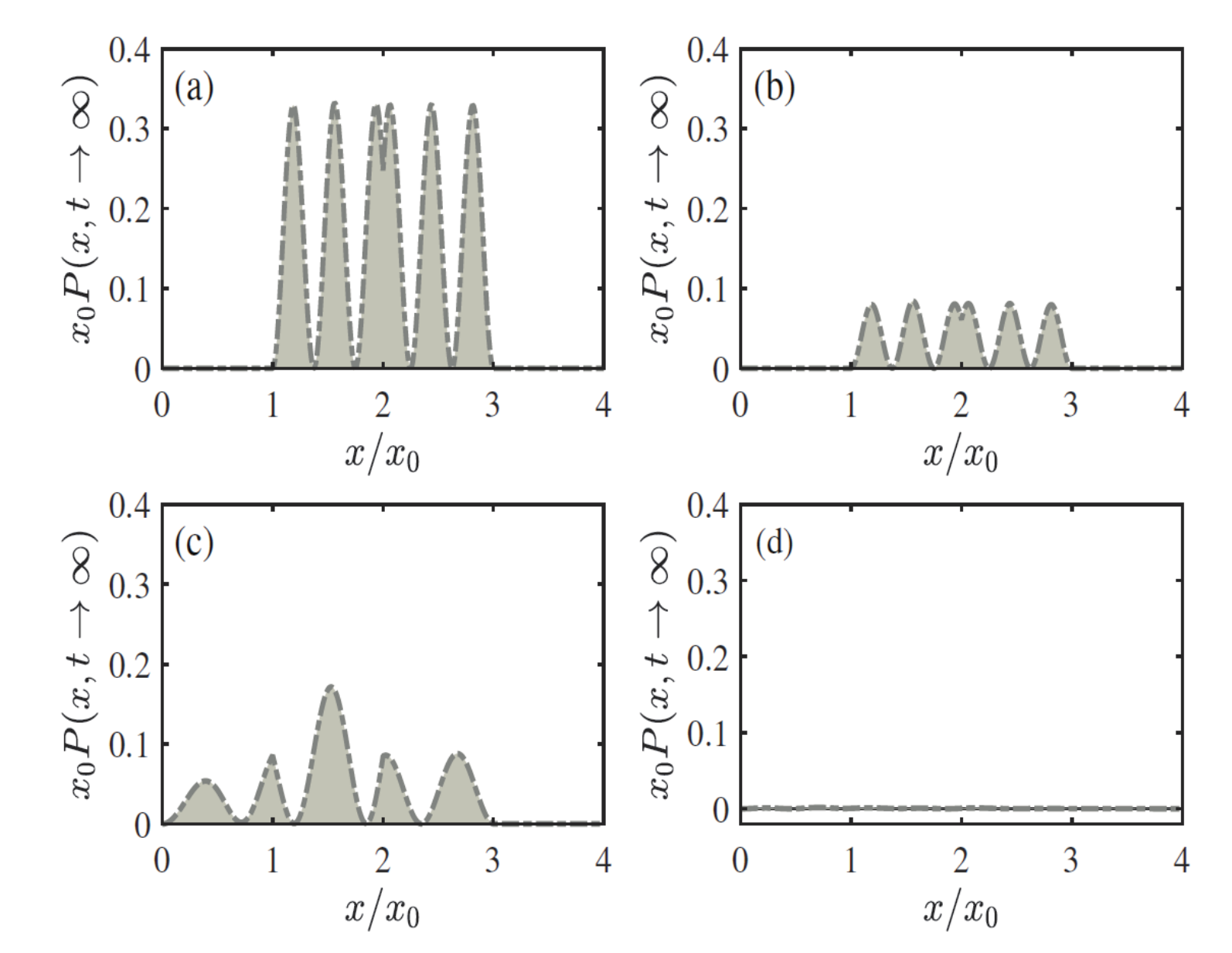}}
\vspace{-0.03cm}
\caption{(Color online) The field intensity distribution $P(x,t)$ in the photonic waveguide at $t\rightarrow \infty$. The four situations shown by Fig.~\ref{3Dcrosssection}(a)-(d) correspond to Fig.~\ref{3D}(a)-(d). The gray-dotted line is the numerical results based on Eq.~(\ref{phi}). The parameters chosen are the same as those in Fig.~\ref{3D}.}\label{3Dcrosssection}
\end{figure}

Figure~\ref{3Dcrosssection} is the long-time field intensity distribution corresponding to Fig.~\ref{3D}, where the orange-dashed line denotes the numerical simulations given by Eq.~(\ref{phi}). In Fig.~\ref{3Dcrosssection}, the field intensities in the above two cases are distributed between the last coupling point and the mirror of the semi-infinite photonic waveguide with $t\rightarrow \infty$.

\begin{figure}[t]
\vspace{0.3cm}
\centerline{
\includegraphics[width=8.8cm, height=6.5cm, clip]{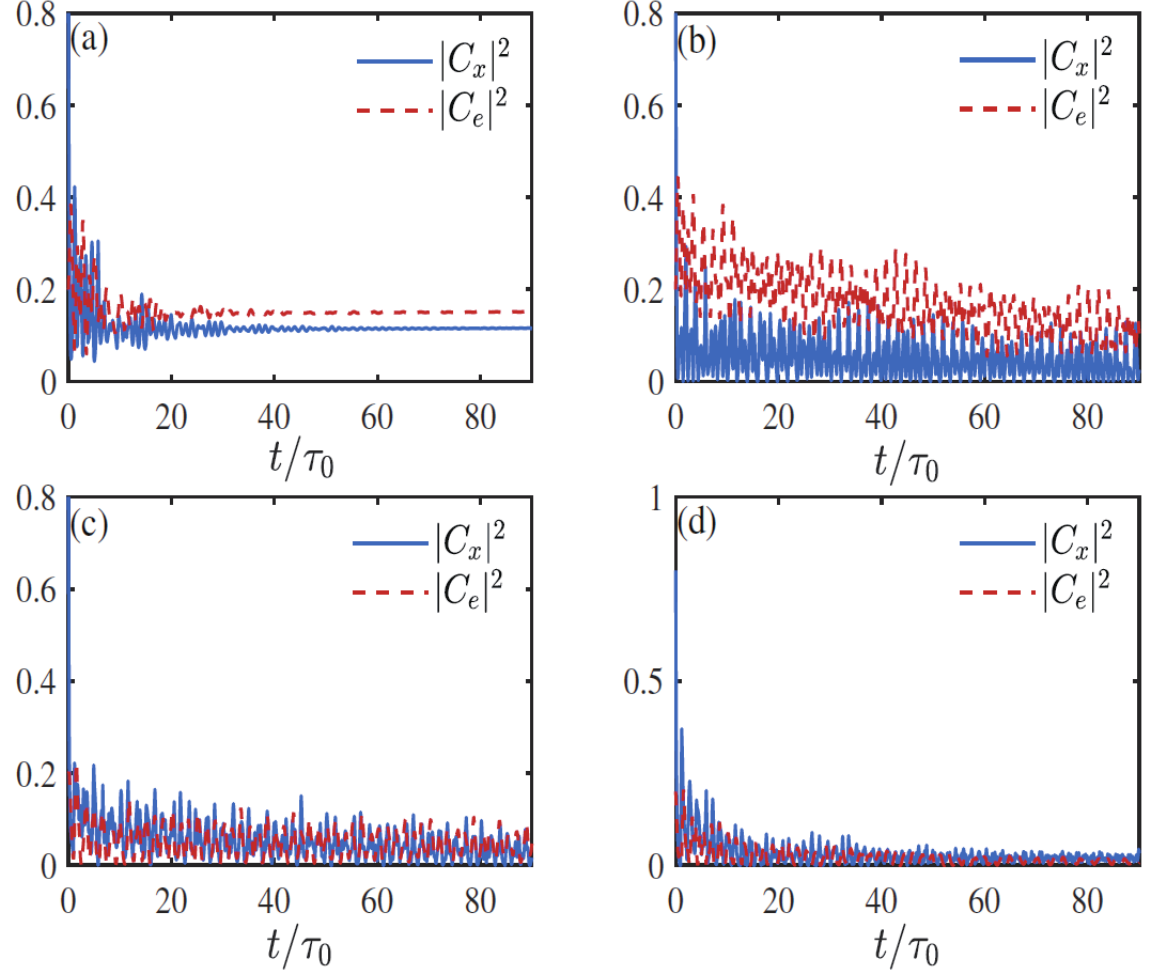}}
\caption{(Color online) Static bound states for the three-level giant atom when the number of coupling points $N$ is different. The parameter $j=6$ is decided by Eq.~(\ref{omegak1}). (a) $N=4$ and $\eta=0.0909$; (b) $N=5$ and $\eta=0.1818$; (c) $N=7$ and $\eta=0.4935$; (d) $N=8$ and $\eta=0.5909$. The red-dashed line represents the numerical solutions in Eq.~(\ref{dotcx}), while the blue-solid line corresponds to the numerical simulation with Eq.~(\ref{dotce}). The other parameters chosen are $G\tau_0=0.7\pi$, $\omega_{x}\tau_{0}=2.2\pi$, $\omega_{e}\tau_{0}=1.1875\pi$, $\omega_{l}\tau_{0}=2.2\pi$, $\Gamma\tau_{0}=0.3\pi$, ${C_x}(0) = \sqrt {0.8} $, and ${C_e}(0) = \sqrt {0.2} $.}\label{case4578}
\end{figure}

Figure~\ref{case4578} plots the atomic excitation probabilities $|C_{x}(t)|^2$ and $|C_{e}(t)|^2$ as functions of time $t$ for varying numbers of coupling points $N$. As $\eta=|\omega_j-\omega_{x}|/\omega_{x}$ increases, the RWA in Fig.~\ref{case4578}(c) and~(d) is no longer satisfied (violating Eq.~(\ref{RWA1})).\\ \\

\subsection{Periodic equal-amplitude oscillating bound states}
For the situation (ii) of Sec. \ref{Sec4}, there is possibility of existing two mode integers $j_{1}$ and $j_{2}$ satisfying the bound-state conditions in Eqs.~(\ref{omegak1}) and (\ref{omegak2}) simultaneously for the same system parameter setting. This is because the bound-state condition for a non-Markovian giant atom given by Eqs.~(\ref{omegak1}) or (\ref{omegak2}) is nonlinear with respect to the mode integer $j$. This means that, in the long-time limit after all the dissipative modes vanish, the atomic excitation probability amplitudes are a superposition of two bound states with different frequencies
$\omega_{j_1}$ and $\omega_{j_2}$. If $j_1=23$ and $j_2=26$ chosen from Fig.~\ref{y1y2}(c) are the two simultaneous solutions of Eq.~(\ref{omegak1}), the parameters $\omega_{x}\tau_0$ and $\Gamma\tau_0$ have to be
\begin{widetext}\begin{align} \begin{cases}
\omega_{x}\tau_0&=\frac{\tau_0G^2}{\omega_{e}+\omega_{l}-\frac{2j_1\pi}{N\tau_0}}
-\cot\left(\frac{j_1\pi}{N}\right)\frac{\left(\frac{\tau_0G^2}{\omega_{e}+\omega_{l}-\frac{2j_1\pi}{N\tau_0}}-\frac{\tau_0G^2}{\omega_{e}+\omega_{l}-\frac{2j_2\pi}{N\tau_0}}+\frac{2j_1\pi}{N}-\frac{2j_2\pi}{N}\right)}{\cot\left(\frac{j_1\pi}{N}\right)-\cot\left(\frac{j_2\pi}{N}\right)}+\frac{2j_1\pi}{N}>0,
\\
\Gamma\tau_0&=\frac{\frac{2}{N}\left(\frac{\tau_0G^2}{\omega_{e}+\omega_{l}-\frac{2j_1\pi}{N\tau_0}}-\frac{\tau_0G^2}{\omega_{e}+\omega_{l}-\frac{2j_2\pi}{N\tau_0}}+\frac{2j_1\pi}{N}-\frac{2j_2\pi}{N}\right)}{\cot\left(\frac{j_1\pi}{N}\right)-\cot\left(\frac{j_2\pi}{N}\right)}>0.
\label{omegakv2}\end{cases}\end{align}
If $j_1=27$ and $j_2=30$ chosen from Fig.~\ref{y1y2}(d) are simultaneously two solutions of Eq.~(\ref{omegak2}), the parameters $\omega_{x}\tau_0$ and $\Gamma\tau_0$ have to be
\begin{align} \begin{cases}
\omega_{x}\tau_0&=\frac{\tau_0G^2}{\omega_{e}+\omega_{l}-\frac{2j_1\pi}{\left(N+1\right)\tau_0}}-\cot\left(\frac{j_1\pi}{N+1}\right)\frac{\left[\frac{\tau_0G^2}{\omega_{e}+\omega_{l}-\frac{2j_1\pi}{\left(N+1\right)\tau_0}}-\frac{\tau_0G^2}{\omega_{e}+\omega_{l}-\frac{2j_2\pi}{\left(N+1\right)\tau_0}}+\frac{2j_1\pi}{N+1}-\frac{2j_2\pi}{N+1}\right]}{\cot\left(\frac{j_1\pi}{N+1}\right)-\cot\left(\frac{j_2\pi}{N+1}\right)}+\frac{2j_1\pi}{N+1}>0,
\\
\Gamma\tau_0&=\frac{\frac{2}{N+1}\left[\frac{\tau_0G^2}{\omega_{e}+\omega_{l}-\frac{2j_1\pi}{\left(N+1\right)\tau_0}}-\frac{\tau_0G^2}{\omega_{e}+\omega_{l}-\frac{2j_2\pi}{\left(N+1\right)\tau_0}}+\frac{2j_1\pi}{N+1}-\frac{2j_2\pi}{N+1}\right]}{\cot\left(\frac{j_1\pi}{N+1}\right)-\cot\left(\frac{j_2\pi}{N+1}\right)}>0.
\label{omegakv3}\end{cases}\end{align}\end{widetext}

Here, $\omega_x \tau_0$ and $\Gamma \tau_0$ are positive values. Moreover, the RWA condition also needs to be met.
To ensure the validity of the RWA, we need $\eta=|\omega_j-\omega_{x}|/\omega_{x}\ll1$. When $\omega_j=2j\pi/N\tau_0-\omega_l/2$ in Eq.~(\ref{skN}), bringing Eq.~(\ref{omegak1}) into $\eta$ gives

\begin{align}
|[\frac12N\Gamma\cot\left(\frac{j\pi}{N}\right)-\frac{G^2}{\omega_{e}+\omega_{l}-\frac{2j\pi}{N\tau_0}}-\frac{\omega_{l}}{2}]/\omega_{x}|\ll1
\label{RWA1},
\end{align}
while for $\omega_j=2j\pi/(N+1)\tau_0-\omega_l/2$ in Eq.~(\ref{skNz1}), substituting Eq.~(\ref{omegak2}) into $\eta$ leads to
\begin{align}
&|[\frac12(N+1)\Gamma\cot\left(\frac{j\pi}{N+1}\right)-\frac{G^2}{\omega_{e}+\omega_{l}-\frac{2j\pi}{(N+1)\tau_0}}-\frac{\omega_{l}}{2}]\nonumber\\
&/\omega_{x}|\ll1.
\label{RWA}
\end{align}

The atomic excitation probability is the superposition of the bound states for frequencies with different modes $j_1$ and $j_2$, while substituting $s_{j_n}(n=1,2)$ in Eq.~(\ref{skN}) into Eqs.~(\ref{Cxtsk}) and~(\ref{Cetsk}), the long-time atomic excitation probability amplitudes read (corresponding to Eq.~(\ref{omegak1}))
\begin{widetext}\begin{align}
C_{x}(t)\approx&\frac{[(-i\frac{2j_{1}\pi}{N\tau_0}+i\frac{\omega_{l}}{2}+i\mu_2)C_{x}(0)-iGC_{e}(0)] e^{-i(\frac{2j_{1}\pi}{N\tau_0}-\frac{\omega_{l}}{2})t}}{[1+\frac{\Gamma}{2}\frac{N\tau_0}{\sin^2(j_{1}\pi/N)}](-i\frac{2j_{1}\pi}{N\tau_0}+i\frac{\omega_{l}}{2}+i\mu_2)+[-i\frac{2j_{1}\pi}{N\tau_0}+i\frac{\omega_{l}}{2}+i\mu_1+\frac{\Gamma}{2}(\frac{2N}{1-e^{i2j_{1}\pi/N}}-N)]}
\notag\\&+\frac{[(-i\frac{2j_{2}\pi}{N\tau_0}+i\frac{\omega_{l}}{2}+i\mu_2)C_{x}(0)-iGC_{e}(0)] e^{-i(\frac{2j_{2}\pi}{N\tau_0}-\frac{\omega_{l}}{2})t}}{[1+\frac{\Gamma}{2}\frac{N\tau_0}{\sin^2(j_{2}\pi/N)}](-i\frac{2j_{2}\pi}{N\tau_0}+i\frac{\omega_{l}}{2}+i\mu_2)+[-i\frac{2j_{2}\pi}{N\tau_0}+i\frac{\omega_{l}}{2}+i\mu_1+\frac{\Gamma}{2}(\frac{2N}{1-e^{i2j_{2}\pi/N}}-N)]}
\label{2CxAnaN},
\\
C_{e}(t)\approx&\frac{\{[-i\frac{2j_{1}\pi}{N\tau_0}+i\frac{\omega_{l}}{2}+i\mu_1+\frac{\Gamma}{2}(\frac{2N}{1-e^{i2j_{1}\pi/N}}-N)] C_{e}(0)-iGC_{x}(0)\} e^{-i(\frac{2j_{1}\pi}{N\tau_0}-\frac{\omega_{l}}{2})t}}{[1+\frac{\Gamma}{2}\frac{N\tau_0}{\sin^2(j_{1}\pi/N)}](-i\frac{2j_{1}\pi}{N\tau_0}+i\frac{\omega_{l}}{2}+i\mu_2)+[-i\frac{2j_{1}\pi}{N\tau_0}+i\frac{\omega_{l}}{2}+i\mu_1+\frac{\Gamma}{2}(\frac{2N}{1-e^{i2j_{1}\pi/N}}-N)]}
\notag\\&+\frac{\{[-i\frac{2j_{2}\pi}{N\tau_0}+i\frac{\omega_{l}}{2}+i\mu_1+\frac{\Gamma}{2}(\frac{2N}{1-e^{i2j_{2}\pi/N}}-N)] C_{e}(0)-iGC_{x}(0)\} e^{-i(\frac{2j_{2}\pi}{N\tau_0}-\frac{\omega_{l}}{2})t}}{[1+\frac{\Gamma}{2}\frac{N\tau_0}{\sin^2(j_{2}\pi/N)}](-i\frac{2j_{2}\pi}{N\tau_0}+i\frac{\omega_{l}}{2}+i\mu_2)+[-i\frac{2j_{2}\pi}{N\tau_0}+i\frac{\omega_{l}}{2}+i\mu_1+\frac{\Gamma}{2}(\frac{2N}{1-e^{i2j_{2}\pi/N}}-N)]}
\label{2CeAnaN}.
\end{align}
Similarly, with Eq.~(\ref{skNz1}), the long-time atomic excitation probability amplitudes corresponding to Eq.~(\ref{omegak2}) are given by
\begin{align}
C_{x}(t)\approx&\frac{\{[-i\frac{2j_{1}\pi}{(N+1)\tau_0}+i\frac{\omega_{l}}{2}+i\mu_2]C_{x}(0)-iGC_{e}(0)\} e^{-i[\frac{2j_{1}\pi}{(N+1)\tau_0}-\frac{\omega_{l}}{2}]t}}{\{1+\frac{\Gamma}{2}\frac{(N+1)\tau_0}{\sin^2[j_{1}\pi/(N+1)]}\}[-i\frac{2j_{1}\pi}{(N+1)\tau_0}+i\frac{\omega_{l}}{2}+i\mu_2]+\{-i\frac{2j_{1}\pi}{(N+1)\tau_0}+i\frac{\omega_{l}}{2}+i\mu_1+\frac{\Gamma}{2}[\frac{2(N+1)}{1-e^{i2j_{1}\pi/(N+1)}}-N-1]\}}
\notag\\&+\frac{\{[-i\frac{2j_{2}\pi}{(N+1)\tau_0}+i\frac{\omega_{l}}{2}+i\mu_2]C_{x}(0)-iGC_{e}(0)\} e^{-i[\frac{2j_{2}\pi}{(N+1)\tau_0}-\frac{\omega_{l}}{2}]t}}{\{1+\frac{\Gamma}{2}\frac{(N+1)\tau_0}{\sin^2[j_{2}\pi/(N+1)]}\}[-i\frac{2j_{2}\pi}{(N+1)\tau_0}+i\frac{\omega_{l}}{2}+i\mu_2]+\{-i\frac{2j_{2}\pi}{(N+1)\tau_0}+i\frac{\omega_{l}}{2}+i\mu_1+\frac{\Gamma}{2}[\frac{2(N+1)}{1-e^{i2j_{2}\pi/(N+1)}}-N-1]\}}
,\label{2CxAnaN1}\\
C_{e}(t)\approx&\frac{\{[-i\frac{2j_{1}\pi}{(N+1)\tau_0}+i\frac{\omega_{l}}{2}+i\mu_1+\frac{\Gamma}{2}\frac{2N+2}{1-e^{i2j_{1}\pi/(N+1)}}-\frac{\Gamma}{2}N-\frac{\Gamma}{2}] C_{e}(0)-iGC_{x}(0)\} e^{-i[\frac{2j_{1}\pi}{(N+1)\tau_0}-\frac{\omega_{l}}{2}]t}}{\{1+\frac{\Gamma}{2}\frac{(N+1)\tau_0}{\sin^2[j_{1}\pi/(N+1)]}\}[-i\frac{2j_{1}\pi}{(N+1)\tau_0}+i\frac{\omega_{l}}{2}+i\mu_2]+\{-i\frac{2j_{1}\pi}{(N+1)\tau_0}+i\frac{\omega_{l}}{2}+i\mu_1+\frac{\Gamma}{2}[\frac{2(N+1)}{1-e^{i2j_{1}\pi/(N+1)}}-N-1]\}}
\notag\\&+\frac{\{[-i\frac{2j_{2}\pi}{(N+1)\tau_0}+i\frac{\omega_{l}}{2}+i\mu_1+\frac{\Gamma}{2}\frac{2N+2}{1-e^{i2j_{2}\pi/(N+1)}}-\frac{\Gamma}{2}N-\frac{\Gamma}{2}] C_{e}(0)-iGC_{x}(0)\} e^{-i[\frac{2j_{2}\pi}{(N+1)\tau_0}-\frac{\omega_{l}}{2}]t}}{\{1+\frac{\Gamma}{2}\frac{(N+1)\tau_0}{\sin^2[j_{2}\pi/(N+1)]}\}[-i\frac{2j_{2}\pi}{(N+1)\tau_0}+i\frac{\omega_{l}}{2}+i\mu_2]+\{-i\frac{2j_{2}\pi}{(N+1)\tau_0}+i\frac{\omega_{l}}{2}+i\mu_1+\frac{\Gamma}{2}[\frac{2(N+1)}{1-e^{i2j_{2}\pi/(N+1)}}-N-1]\}}
\label{2CeAnaN1},
\end{align}\end{widetext}
where $j_1$ and $j_2$ are two solutions satisfying the transcendental equations in Eqs.~(\ref{omegak1}) and (\ref{omegak2}) (see Fig.~\ref{y1y2}(c) and (d)).
\begin{figure}[t]
\centerline{
\includegraphics[width=8.9cm, height=7.0cm, clip]{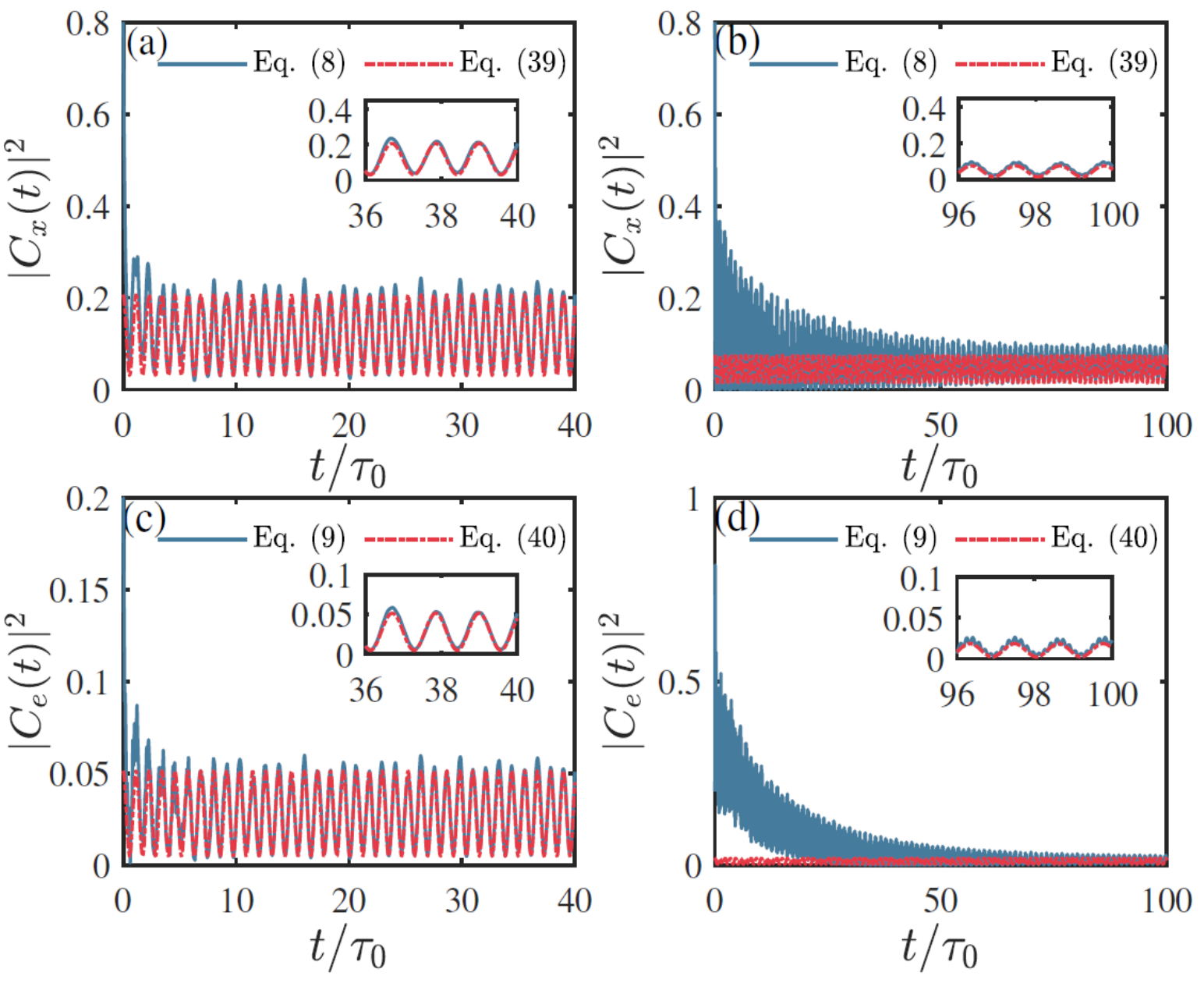}}
\vspace{-0.4cm}
\caption{(Color online) Periodic equal-amplitude oscillating bound states in the photonic waveguide for the three-level giant atom with the number of coupling points $N = 6$. $j_1=23$ and $j_2=26$ are chosen from Fig.~\ref{y1y2}(c). The red-dashed line represents the analytical solutions in Eqs.~(\ref{2CxAnaN}) and~(\ref{2CeAnaN}), while the blue-solid line denotes the numerical simulation with Eqs.~(\ref{dotcx}) and (\ref{dotce}), respectively. In Fig.~\ref{TwoSolution1}(a) and~(c), we take $G\tau_0=2.8\pi$. In Fig.~\ref{TwoSolution1}(b) and~(d), we take $G\tau_0=-2.8\pi$. The insets in the figure magnify the probability amplitudes. The other parameters chosen are $\omega_x \tau_0=7.0366\pi$, $\omega_e \tau_0=0.2 \pi$, $\omega_l \tau_0=2.5 \pi$, $\Gamma \tau_0=0.1825\pi$, ${C_x}(0) = \sqrt {0.8} $, and ${C_e}(0) = \sqrt {0.2} $.}\label{TwoSolution1}
\end{figure}
\begin{figure}[t]
\centerline{
\includegraphics[width=8.9cm, height=7.0cm, clip]{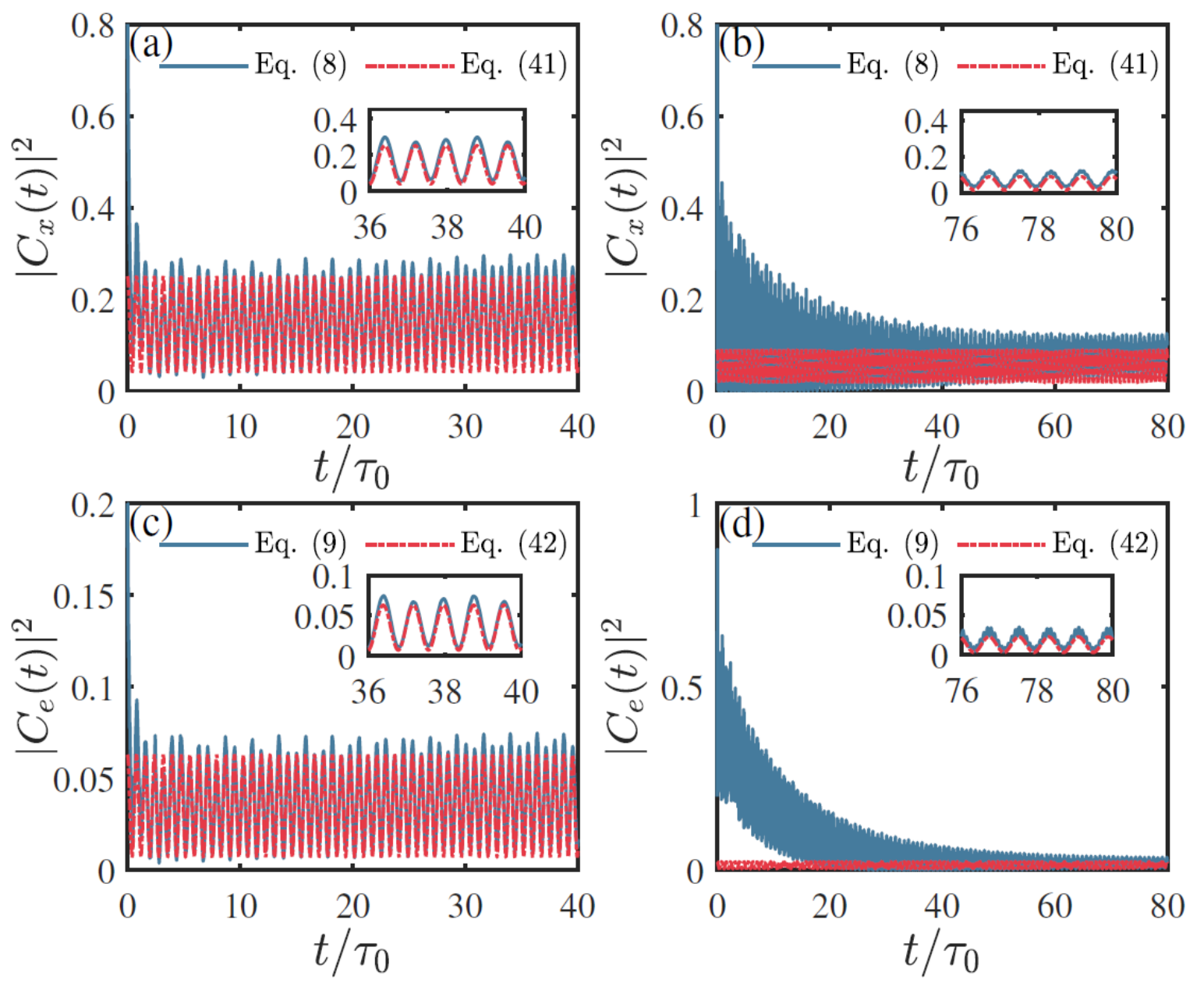}}
\vspace{-0.4cm}
\caption{(Color online) Periodic equal-amplitude oscillating bound states in the photonic waveguide for the three-level giant atom with the number of coupling points $N = 6$. $j_1=27$ and $j_2=30$ are chosen from Fig.~\ref{y1y2}(d). The red-dashed line denotes the analytical solutions in Eqs.~(\ref{2CxAnaN1}) and~(\ref{2CeAnaN1}), while the blue-solid line corresponds to the numerical simulation with Eqs.~(\ref{dotcx}) and (\ref{dotce}), respectively. In Fig.~\ref{TwoSolution2}(a) and~(c), we take $G\tau_0=2.8\pi$. In Fig.~\ref{TwoSolution2}(b) and~(d), we take $G\tau_0=-2.8\pi$. The insets in the figure magnify the probability amplitudes. The other parameters chosen are $\omega_x \tau_0=6.9350\pi$, $\omega_e \tau_0=0.2 \pi$, $\omega_l \tau_0=2.5 \pi$, $\Gamma \tau_0=0.1079\pi$, ${C_x}(0) = \sqrt {0.8}$, and ${C_e}(0) = \sqrt{0.2} $.}\label{TwoSolution2}
\end{figure}
\begin{figure}[t]
\centerline{
\includegraphics[width=8.9cm, height=6.3cm, clip]{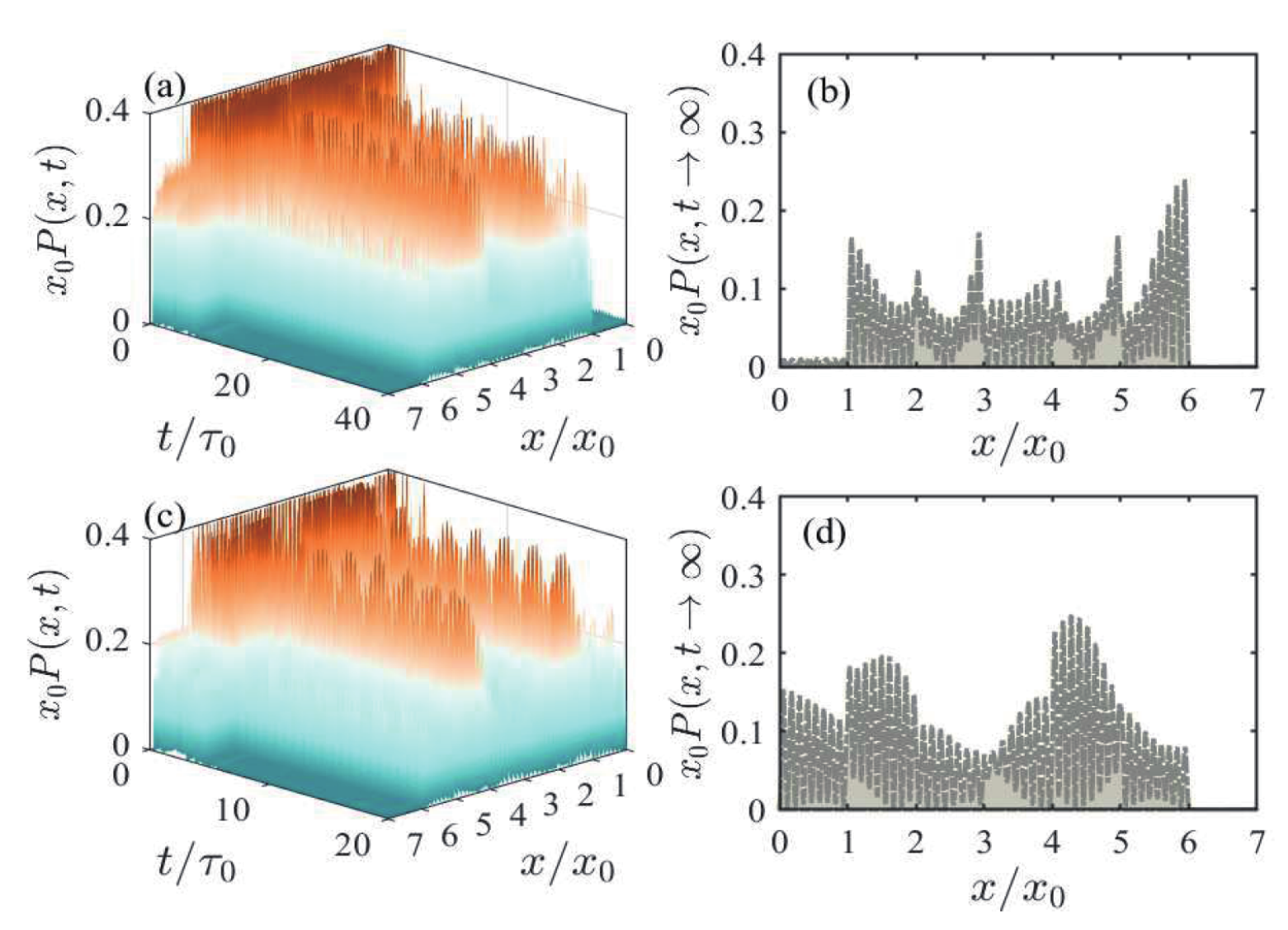}}
\vspace{-0.01cm}
\caption{(Color online) Figure~\ref{PEA3d}(a) and (c) represent the time evolution of field intensity in the photonic waveguide with the number of coupling points $N = 6$. Figure~\ref{PEA3d}(b) and (d) denote the corresponding field intensity distribution $P(x,t)$ in the semi-infinite photonic waveguide at $t\rightarrow \infty$. The top line is plotted by the bound state condition in Eq.~(\ref{omegak1}), where the corresponding parameters are the same as Fig.~\ref{TwoSolution1}(a) and~(c). The bottom line is plotted by the bound state condition in Eq.~(\ref{omegak2}), where the parameters chosen are the same as those in Fig.~\ref{TwoSolution2}(a) and (c).}\label{PEA3d}
\end{figure}

When the driving strength $G=0$ and driving frequency $\omega_l=0$, the long-time dynamics in Eqs.~(\ref{2CxAnaN}) and (\ref{2CeAnaN}) is consistent with that given in Ref.\cite{Johansson2020}. However, another solution with the term of the factor $N+1$ in Eqs.~(\ref{2CxAnaN1}) and (\ref{2CeAnaN1}) differs from that in Ref.\cite{Johansson2020}. The atomic excitation probability for a six-coupling-points giant atom ($N = 6$) with two coexisting bound states is shown in Figs.~\ref{TwoSolution1} and~\ref{TwoSolution2} (The values of $j_1=23$ and $j_2=26$ are the same as Fig.~\ref{y1y2}(c) in Fig.~\ref{TwoSolution1}, while the values of $j_1=27$ and $j_2=30$ are the same as Fig.~\ref{y1y2}(d) in Fig.~\ref{TwoSolution2}). In Fig.~\ref{TwoSolution1}, we choose the values of the two mode integers to be $j_{1}=23$ and $j_{2}=26$ according to Eq.~(\ref{omegakv2}). In Fig.~\ref{TwoSolution2}, we choose the values of the two mode integers to be $j_{1}=27$ and $j_{2}=30$ according to Eq.~(\ref{omegakv3}). The red-dashed line corresponds to the analytical solutions in Eqs.~(\ref{2CxAnaN}),~(\ref{2CeAnaN}),~(\ref{2CxAnaN1}), and~(\ref{2CeAnaN1}), while the blue-solid line represents the numerical simulation with Eqs.~(\ref{dotcx}) and~(\ref{dotce}), respectively. In Fig.~\ref{PEA3d}(a) and~(c), we plot the corresponding time evolution of the field intensity function $P(x,t)$ at position $x$ in the photonic waveguide, showing a periodic equal-amplitude oscillating bound state in the long-time limit. In Fig.~\ref{PEA3d}(b) and~(d), we show the long-time field intensity distribution calculated by Eq.~(\ref{phi}) corresponding to Fig.~\ref{PEA3d}(a) and~(b).

\section{infinite-photonic waveguide case}\label{Sec7}
\begin{figure}[t]
\centerline{
\includegraphics[width=8.6cm, height=3.55cm, clip]{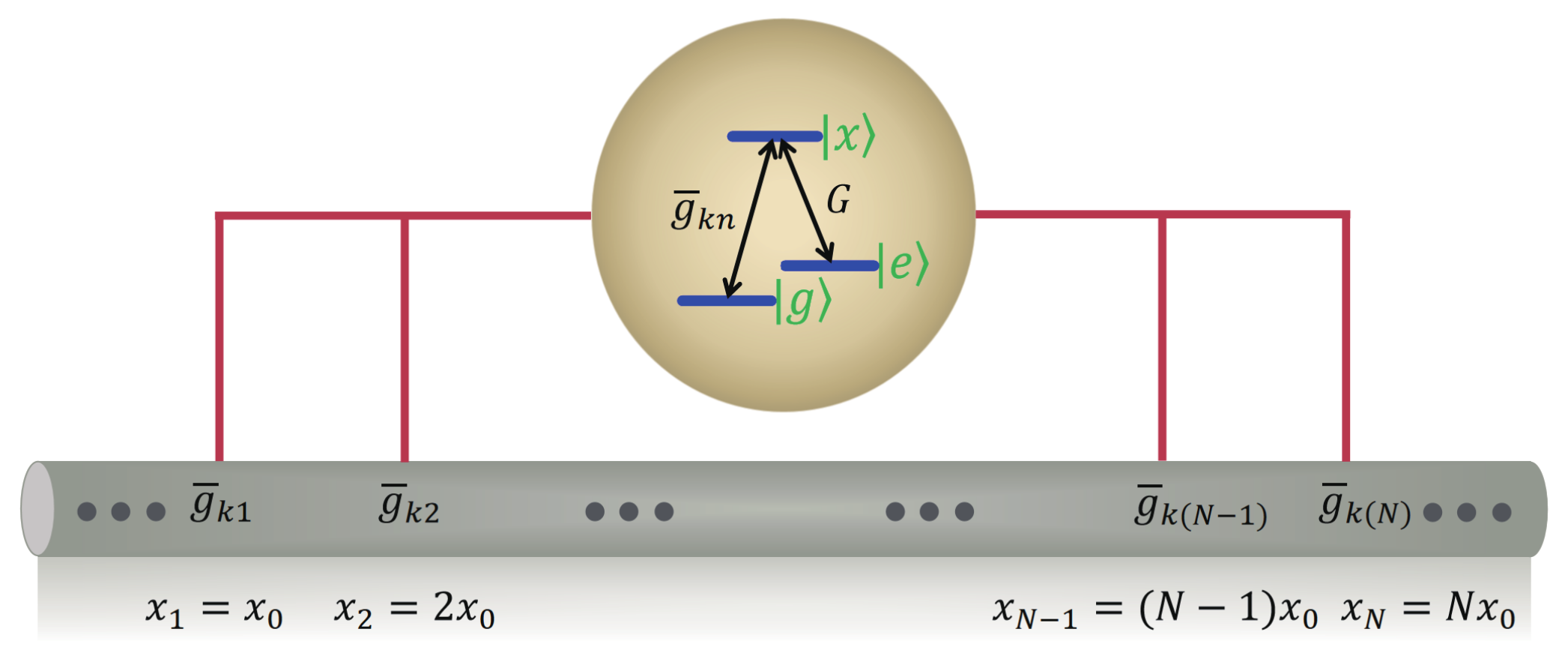}}
\vspace{-0.3cm}
\caption{(Color online) This figure shows that the $\Lambda$-type three-level giant atom driven by a classical driving field couples with a one-dimensional infinite photonic waveguide through $N$ coupling points. The transition between the ground state $|g\rangle$ and the excited state $|x\rangle$ is coupled to a photonic waveguide positioned at $x_n$ with the coupling coefficient $\bar{g}_{kn}=\sqrt{\Gamma v/2\pi}e^{iknx_0}$. The distance between the adjacent coupling points is $x_0$. The laser field with the frequency $\omega_l$ and driving strength $G$ drives the transition between the excited state $|x\rangle$ and the metastable state $|e\rangle$.}\label{model2}
\end{figure}

In this section, we study the non-Markovian dynamics for the driven three-level giant atom coupling with a one-dimensional infinite photonic waveguide, which is shown in Fig.~\ref{model2}. Under the RWA, we write the total system Hamiltonian of the case as
\begin{equation}\begin{aligned}
{\hat H}_{inf}=&\mu_1\hat{\sigma}_{xx}+\mu_2\hat{\sigma}_{ee}\\
&+\int(\Omega_{k}-\frac{\omega_{l}}{2})\hat a_{k}^{\dagger}\hat a_{k}dk+G(\hat{\sigma}_{ex}+\hat{\sigma}_{xe})\\
&+\sum_{n=1}^{N}\int dk(\bar{g}_{kn}\hat a_{k}^{\dagger}\hat{\sigma}_{gx}+\bar{g}_{kn}^{*}\hat a_{k}\hat{\sigma}_{xg}),
\label{hatHinf}\end{aligned}\end{equation}
where $\mu_1=\omega_{x}-\frac{\omega_{l}}{2}$ and $\mu_2=\omega_{e}+\frac{\omega_{l}}{2}$. The coupling coefficient between the giant atom and the infinite photonic waveguide is changed to $\bar{g}_{kn}=\sqrt{\Gamma v/2\pi}e^{iknx_0}$. Since the RWA guarantees that there is only one excitation either in the atom or in the photonic waveguide when the total system initially has one excitation, we study the non-Markovian dynamics of giant atom in the single-excitation subspace of the total system. Since the total system is also initially prepared in state $\left|{{\bar{\psi}(0)}}\right\rangle=\bar{C}_x(0)\left|{x,0}\right\rangle+\bar{C}_e(0)\left|{e,0}\right\rangle$, the description of the state for the system at time $t$ is
\begin{equation}\begin{aligned}
\left|{{\bar{\psi}(t)}}\right\rangle=\bar{C}_x(t)\left|{x,0}\right\rangle+\bar{C}_e(t)\left|{e,0}\right\rangle+\int{dk\bar{\beta} _k(t)} \left|{g,{1_k}}\right\rangle,
\label{psitinf}\end{aligned}\end{equation}
where $\bar{C}_{x}(t)$, $\bar{C}_{e}(t)$, and $\bar{\beta}_{k}(t)$ denote the probability amplitudes corresponding to states $\left| {x,0} \right\rangle$, $\left| {e,0} \right\rangle$, and $\left| {g,1_k} \right\rangle$, respectively.

Compared with the infinite photonic waveguide, the case of the semi-infinite photonic waveguide given by Eq.~(\ref{dotcx}) has an additional feedback term represented by $\frac{\Gamma}{2}\sum_{m,n=1}^{N}\bar{C}_{x}[t-(m+n)\tau_0]e^{i\frac{\omega_{l}}{2}(m+n)\tau_0}\Theta[t-(m+n)\tau_0]$, which originates from the presence of a mirror at the endpoint of the semi-infinite photonic waveguide. This feedback term disappears when there is no mirror, where the system is equivalent to a three-level giant atom coupling with a one-dimensional infinite photonic waveguide. In this case, Eqs.~(\ref{dotcx}) and~(\ref{dotce}) are modified as follows (the calculation process is performed in Appendix \ref{D})

\begin{align}
\dot{\bar{C}}_{x}(t)=&-i\mu_1 \bar{C}_{x}(t)-iG\bar{C}_{e}(t) -\frac{\Gamma}{2}\sum_{m,n=1}^{N} e^{i\frac{\omega_{l}}{2}|m-n|\tau_0}\nonumber\\
&\cdot \bar{C}_{x}(t-|m-n|\tau_0)\Theta(t-\left|m-n\right|\tau_0),
\label{dotcxinf}\\
\dot{\bar{C}}_{e}(t)=&-i\mu_2 \bar{C}_{e}(t)-iG\bar{C}_{x}(t),\label{dotceinf}
\end{align}
which can return back to results in Ref.\cite{Johansson2020} when $G=\omega_l=\omega_e=0$ and also be reduced to the case in Ref.\cite{Xin2022} with three-level point-like atom at $N=1$. Thus, the probability amplitude $\bar{\phi}\left(x,t\right)$ of the real space field in the infinite photonic waveguide is given by

\begin{align}
\bar{\phi}\left(x,t\right)=&-i\sqrt{\frac{\Gamma}{2v}}\sum_{n=1}^{N}\bar{C}_{x}(t-|\tau-n\tau_0|)\nonumber\\
&\cdot e^{i\frac{\omega_{l}}{2}|\tau-n\tau_0|}\Theta(t-|\tau-n\tau_0|).\nonumber\\
\label{phiinf}
\end{align}

In order to solve Eqs.~(\ref{dotcxinf}) and~(\ref{dotceinf}), we use Laplace transformation and obtain
\begin{small}
\begin{eqnarray}
s\tilde{\bar{C}}_{x}\left(s)-\bar{C}_{x}(0\right) &=&-i\mu_1\tilde{\bar{C}}_{x}(s)-i G\tilde{\bar{C}}_{e}\left(s\right)\nonumber\\
&&
-\frac{\Gamma}{2}\sum_{m,n=1}^{N}\tilde{\bar{C}}_{x}(s)e^{-(s-i\omega_{l}/2)\left|m-n\right|\tau_0}
,\label{10inf}\\
s\tilde{\bar{C}}_{e}(s)-\bar{C}_{e}(0)&=&-i\mu_2 \tilde{\bar{C}}_{e}(s)-iG\tilde{\bar{C}}_{x}\left(s\right)
.\label{11inf}
\end{eqnarray}
\end{small}
Therefore, we have
\begin{align}
\tilde{\bar{C}}_{x}\left(s\right)&=\frac{\bar{C}_{x}\left(0)(s+i\mu_2\right)-iG\bar{C}_{e}\left(0\right)}{[s+i\mu_1+\gamma_{1}(s)](s+i\mu_2)+G^2},
\label{tildeCxsinf}
\\
\tilde{\bar{C}}_{e}\left(s\right)&=\frac{\bar{C}_{e}(0)[s+i\mu_1+\gamma_{1}(s)]-iG\bar{C}_{x}\left(0\right)}{[s+i\mu_1+\gamma_{1}(s)](s+i\mu_2)+G^2},
\label{tildeCesinf}
\end{align}
where $\gamma_{1}(s)$ is given by
\begin{small}
\begin{eqnarray}
\gamma_{1}(s)&=&\frac{\Gamma}{2}\sum_{m,n=1}^{N}e^{-(s-i\omega_{l}/2)|m-n|\tau_0}.
\end{eqnarray}
\end{small}
Inverting Laplace transformation to Eqs.~(\ref{dotcxinf}) and~(\ref{dotceinf}) results in
\begin{small}\begin{align}
\bar{C}_{x}(t)&=\sum_{\bar{j}}\frac{[\bar{C}_{x}\left(0)(s_{\bar{j}}+i\mu_2\right)-iG^2\bar{C}e(0)]e^{-i\omega_{\bar{j}}t}}{[1+\gamma_{2}(s_{\bar{j}})](s_{\bar{j}}+i\mu_2)+[s_{\bar{j}}+i\mu_1+\gamma_{1}(s_{\bar{j}})]}\nonumber\\
&+\sum_{\bar{\alpha}}\frac{[\bar{C}_{x}\left(0)(s_{\bar{\alpha}}+i\mu_2\right)-iG^2\bar{C}e(0)]e^{s_{\bar{\alpha}}t}}{[1+\gamma_{2}(s_{\bar{\alpha}})](s_{\bar{\alpha}}+i\mu_2)+[s_{\bar{\alpha}}+i\mu_1+\gamma_{1}(s_{\bar{\alpha}})]},
\label{Cxtskinf}
\\
\bar{C}_{e}(t)&=\sum_{\bar{j}}\frac{\{ \bar{C}_{e}(0)[s_{\bar{j}}+i\mu_1+\gamma_{1}(s_{\bar{j}})]-iG\bar{C}_{x}(0)\}e^{-i\omega_{\bar{j}}t}}{[1+\gamma_{2}(s_{\bar{j}})](s_{\bar{j}}+i\mu_2)+[s_{\bar{j}}+i\mu_1+\gamma_{1}(s_{\bar{j}})]}\nonumber\\
&+\sum_{\bar{\alpha}}\frac{\{ \bar{C}_{e}(0)[s_{\bar{\alpha}}+i\mu_1+\gamma_{1}(s_{\bar{\alpha}})]-iG\bar{C}_{x}(0)\}e^{s_{\bar{\alpha}}t}}{[1+\gamma_{2}(s_{\bar{\alpha}})](s_{\bar{\alpha}}+i\mu_2)+[s_{\bar{\alpha}}+i\mu_1+\gamma_{1}(s_{\bar{\alpha}})]}, \label{Cetskinf}
\end{align}\end{small}
with
\begin{small}
\begin{eqnarray}
\gamma_{2}(s)&=&-\frac{\Gamma}{2}\sum_{m,n=1}^{N}|m-n|\tau_0e^{-(s-i\omega_{l}/2)|m-n|\tau_0},
\end{eqnarray}
\end{small}
where the first term of Eqs.~(\ref{Cxtskinf}) and (\ref{Cetskinf}) corresponds to the contribution of the bound states (oscillating with time), while the second term represents the oscillation damping  (approaching zero  in long-time limit) due to the complex roots $s_{\bar{\alpha}}$.

All poles of $\tilde{\bar{C}}_{x}(s)$ and $\tilde{\bar{C}}_{e}(s)$ are given by all complex roots (including the pure imaginary roots $s_{\bar{j}}=-i\omega_{\bar{j}}$ with the real number $\omega_{\bar{j}}$, see the discussions below Eq.~(\ref{skinf})) of the following equation:
\begin{eqnarray}
&&(s_{\bar{\alpha}}+i\mu_1+\frac{\Gamma}{2}\sum_{m,n=1}^{N}e^{-(s_{\bar{\alpha}}-i\omega_{l}/2)\left|m-n\right|\tau_0}) \nonumber \\
&&\cdot (s_{\bar{\alpha}}+i\mu_2)+G^2=0
.\label{skinf}
\end{eqnarray}

\begin{figure}[t]
\centerline{
\includegraphics[width=8.8cm, height=6.5cm, clip]{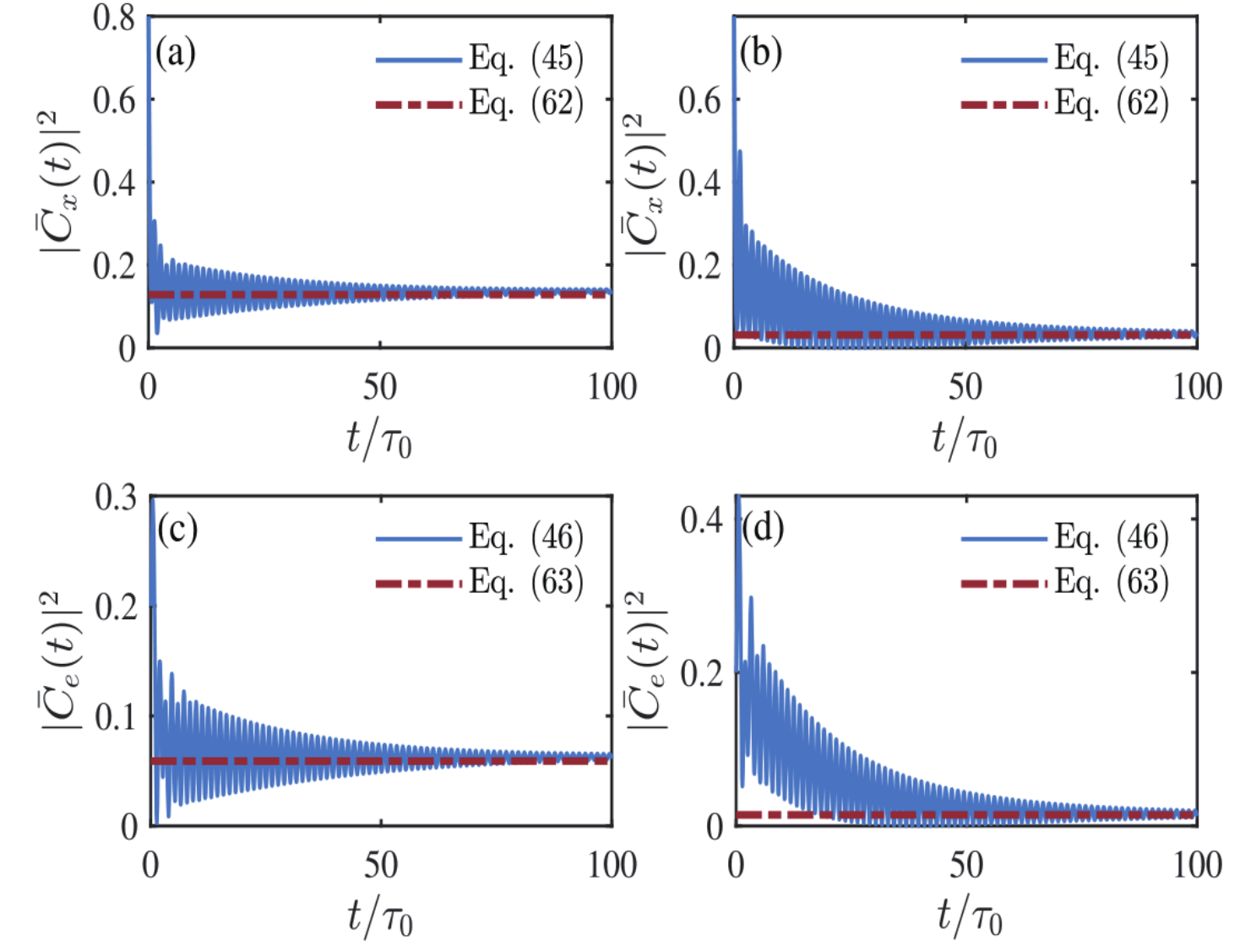}}
\vspace{-0.01cm}
\caption{(Color online) Static bound states for the three-level giant atom in the infinite photonic waveguide with the number of coupling points $N = 3$. The parameter $\bar{j}=4$ is chosen from Fig.~\ref{y1y2}(a). The red-dashed line represents the analytical solutions in Eqs.~(\ref{CxAnaNinf}) and (\ref{CeAnaNinf}), while the blue-solid line corresponds to the numerical simulation with Eqs.~(\ref{dotcxinf}) and (\ref{dotceinf}). In Fig.~\ref{inStaticBS}(a) and (c), we take $G \tau_0= 0.6\pi ¦£$. In Fig.~\ref{inStaticBS}(b) and~(d), we take $G \tau_0= -0.6\pi ¦£$. The other parameters chosen are $\omega_x \tau_0=2\pi$, $\omega_e \tau_0=0.5818\pi$, $\omega_l \tau_0=1.2 \pi$, $\Gamma \tau_0=0.3 \pi$, ${\bar{C}_x}(0) = \sqrt {0.8} $, and ${\bar{C}_e}(0) = \sqrt {0.2} $.}\label{inStaticBS}
\end{figure}

\begin{figure}
\includegraphics[width=8.2cm]{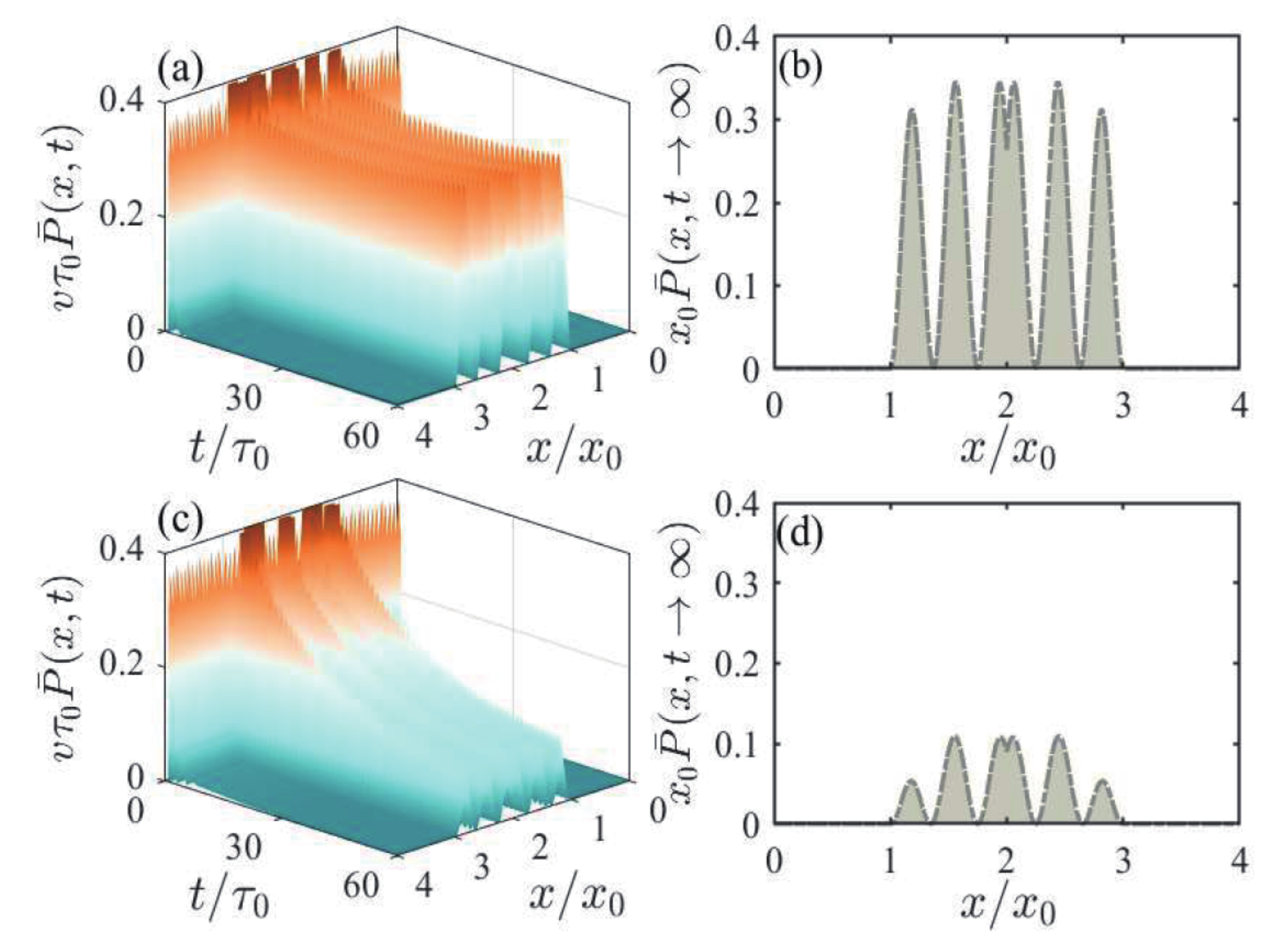}
\vspace{-0.01cm}
\caption{(Color online) This figure is plotted by the bound state condition in Eq.~(\ref{omegak1inf}). Figure~\ref{in3D3Dcross}(a) and (c) denote time evolution of field intensity in the infinite photonic waveguide with the number of coupling points $N = 3$. Figure~\ref{in3D3Dcross}(b) and (d) represent the corresponding field intensity distribution $\bar{P}(x,t)$ in the photonic waveguide at $t\rightarrow \infty$. The driving strength of left column is $G\tau_0=0.6\pi$. The driving strength of right column is $G\tau_0=-0.6\pi$. The parameter $\bar{j}=4$ is obtained by Fig.~\ref{y1y2}(a). The other parameters chosen are $\omega_x \tau_0=2\pi$, $\omega_e \tau_0=0.5818\pi$, $\omega_l \tau_0=1.2 \pi$, $\Gamma \tau_0=0.3 \pi$, ${\bar{C}_x}(0) = \sqrt {0.8} $, and ${\bar{C}_e}(0) = \sqrt {0.2} $.}\label{in3D3Dcross}
\end{figure}

In order to find the pure imaginary solution $s_{\bar{j}}=-i\omega_{\bar{j}}$ ($\omega_{\bar{j}}$ denotes the real number), we divide Eq.~(\ref{skinf}) into imaginary and real parts respectively described through
\begin{small}\begin{align}
&\frac{1}{4}\Gamma\csc^2\left(\frac{1}{2}\tilde{\omega}_{\bar{j}}\tau_0\right) \nonumber\\
&\cdot \left(2\omega_{e}-2\omega_{\bar{j}}+\omega_{l}\right)\sin^2\left(\frac{N\tilde{\omega}_{\bar{j}}\tau_0}{2}\right)=0
,\label{iminf}\\
&\frac{1}{8}(2\omega_{e}-2\omega_{\bar{j}}+\omega_{l})[\Gamma\csc^2\left(\frac{1}{2}\tilde{\omega}_{\bar{j}}\tau_0\right)\sin\left(N\tilde{\omega}_{\bar{j}}\tau_0\right)\nonumber\\
&-2N\Gamma\cot\left(\frac{1}{2}\tilde{\omega}_{\bar{j}}\tau_0\right)+4\omega_{\bar{j}}+2\omega_{l}-4\omega_{x}]+G^2=0
,\label{reinf}
\end{align}\end{small}
where $\tilde{\omega}_{\bar j}=(2\omega_{\bar j}+\omega_{l})/2$. Solving Eqs.~(\ref{iminf}) and (\ref{reinf}) simultaneously, we obtain
\begin{align}
s_{\bar{j}} & =-i\omega_{\bar{j}}=-i\frac{2\bar{j}\pi}{N\tau_0}+i\frac{\omega_{l}}{2}
.\label{skNinf}
\end{align}
Here, we obtain only a purely imaginary solution originating from the absence of mirror at one end of the photonic waveguide (leading to the last term in Eq.~(\ref{dotcx}) disappearing), which is significantly different from the case of the semi-infinite photonic waveguide in Sec.~\ref{Sec2} - Sec.~\ref{Sec5}. This also indicates that the atomic excitation probability can be manipulated through the mirror in the semi-infinite photonic waveguide. Then, the corresponding bound state condition satisfies
\begin{align}
\omega_{x}\tau_0&=\frac{2\bar{j}\pi}{N}-\frac{1}{2}N\Gamma\tau_0\cot\left(\frac{\bar{j}\pi}{N}\right)+\frac{\tau_0G^2}{\omega_{e}+\omega_{l}-\frac{2\bar{j}\pi}{N\tau_0}}
,\label{omegak1inf}
\end{align}
or equivalently
\begin{align}
\omega_{e}\tau_0&=\frac{2\tau_0G^2}{\Gamma N\cot\left(\frac{\bar{j}\pi}{N}\right)-\frac{4\bar{j}\pi}{N\tau_0}+2\omega_{x}}-\omega_{l}\tau_0+\frac{2\bar{j}\pi}{N} .\label{bs1inf}
\end{align}
Substituting $s_{\bar{j}}$ in Eq.~(\ref{skNinf}) into Eqs.~(\ref{Cxtskinf}) and~(\ref{Cetskinf}), the long-time dynamics of the atomic excitation probability amplitudes for this case reads
\begin{widetext}\begin{align}
\bar{C}_{x}(t)&\approx\frac{[(-i\frac{2\bar{j}\pi}{N\tau_0}+i\frac{\omega_{l}}{2}+i\mu_2)\bar{C}_{x}(0)-iG\bar{C}_{e}(0)] e^{-i(\frac{2\bar{j}\pi}{N\tau_0}-\frac{\omega_{l}}{2})t}}{[1+\frac{\Gamma}{2}\frac{N\tau_0}{\sin^2(\bar{j}\pi/N)}](-i\frac{2\bar{j}\pi}{N\tau_0}+i\frac{\omega_{l}}{2}+i\mu_2)+[-i\frac{2\bar{j}\pi}{N\tau_0}+i\frac{\omega_{l}}{2}+i\mu_1+\frac{\Gamma}{2}(\frac{2N}{1-e^{i2\bar{j}\pi/N}}-N)]}
\label{CxAnaNinf},
\\
\bar{C}_{e}(t)&\approx\frac{\{[-i\frac{2\bar{j}\pi}{N\tau_0}+i\frac{\omega_{l}}{2}+i\mu_1+\frac{\Gamma}{2}(\frac{2N}{1-e^{i2\bar{j}\pi/N}}-N)] \bar{C}_{e}(0)-iG\bar{C}_{x}(0)\} e^{-i(\frac{2\bar{j}\pi}{N\tau_0}-\frac{\omega_{l}}{2})t}}{[1+\frac{\Gamma}{2}\frac{N\tau_0}{\sin^2(\bar{j}\pi/N)}](-i\frac{2\bar{j}\pi}{N\tau_0}+i\frac{\omega_{l}}{2}+i\mu_2)+[-i\frac{2\bar{j}\pi}{N\tau_0}+i\frac{\omega_{l}}{2}+i\mu_1+\frac{\Gamma}{2}(\frac{2N}{1-e^{i2\bar{j}\pi/N}}-N)]}
,\label{CeAnaNinf}
\end{align}
where $\bar j$ is one solution satisfying the transcendental equation in Eq.~(\ref{omegak1inf}) (see Fig.~\ref{y1y2}(a)). The conditions to be satisfied for the coexistence of two bound states are consistent with Eq.~(\ref{omegakv2}). We also obtain the long-time dynamics of the atomic excitation probability amplitudes for a three-level giant atom with two coexisting bound states
\begin{align}
\bar{C}_{x}(t)\approx&\frac{[(-i\frac{2\bar{j}_{1}\pi}{N\tau_0}+i\frac{\omega_{l}}{2}+i\mu_2)\bar{C}_{x}(0)-iG\bar{C}_{e}(0)] e^{-i(\frac{2\bar{j}_{1}\pi}{N\tau_0}-\frac{\omega_{l}}{2})t}}{[1+\frac{\Gamma}{2}\frac{N\tau_0}{\sin^2(\bar{j}_{1}\pi/N)}](-i\frac{2\bar{j}_{1}\pi}{N\tau_0}+i\frac{\omega_{l}}{2}+i\mu_2)+[-i\frac{2\bar{j}_{1}\pi}{N\tau_0}+i\frac{\omega_{l}}{2}+i\mu_1+\frac{\Gamma}{2}(\frac{2N}{1-e^{i2\bar{j}_{1}\pi/N}}-N)]}
\notag\\&+\frac{[(-i\frac{2\bar{j}_{2}\pi}{N\tau_0}+i\frac{\omega_{l}}{2}+i\mu_2)\bar{C}_{x}(0)-iG\bar{C}_{e}(0)] e^{-i(\frac{2\bar{j}_{2}\pi}{N\tau_0}-\frac{\omega_{l}}{2})t}}{[1+\frac{\Gamma}{2}\frac{N\tau_0}{\sin^2(\bar{j}_{2}\pi/N)}](-i\frac{2\bar{j}_{2}\pi}{N\tau_0}+i\frac{\omega_{l}}{2}+i\mu_2)+[-i\frac{2\bar{j}_{2}\pi}{N\tau_0}+i\frac{\omega_{l}}{2}+i\mu_1+\frac{\Gamma}{2}(\frac{2N}{1-e^{i2\bar{j}_{2}\pi/N}}-N)]}
\label{2CxAnaNinf},
\\
\bar{C}_{e}(t)\approx&\frac{\{[-i\frac{2\bar{j}_{1}\pi}{N\tau_0}+i\frac{\omega_{l}}{2}+i\mu_1+\frac{\Gamma}{2}(\frac{2N}{1-e^{i2\bar{j}_{1}\pi/N}}-N)] \bar{C}_{e}(0)-iG\bar{C}_{x}(0)\} e^{-i(\frac{2\bar{j}_{1}\pi}{N\tau_0}-\frac{\omega_{l}}{2})t}}{[1+\frac{\Gamma}{2}\frac{N\tau_0}{\sin^2(\bar{j}_{1}\pi/N)}](-i\frac{2\bar{j}_{1}\pi}{N\tau_0}+i\frac{\omega_{l}}{2}+i\mu_2)+[-i\frac{2\bar{j}_{1}\pi}{N\tau_0}+i\frac{\omega_{l}}{2}+i\mu_1+\frac{\Gamma}{2}(\frac{2N}{1-e^{i2\bar{j}_{1}\pi/N}}-N)]}
\notag\\&+\frac{\{[-i\frac{2\bar{j}_{2}\pi}{N\tau_0}+i\frac{\omega_{l}}{2}+i\mu_1+\frac{\Gamma}{2}(\frac{2N}{1-e^{i2\bar{j}_{2}\pi/N}}-N)] \bar{C}_{e}(0)-iG\bar{C}_{x}(0)\} e^{-i(\frac{2\bar{j}_{2}\pi}{N\tau_0}-\frac{\omega_{l}}{2})t}}{[1+\frac{\Gamma}{2}\frac{N\tau_0}{\sin^2(\bar{j}_{2}\pi/N)}](-i\frac{2\bar{j}_{2}\pi}{N\tau_0}+i\frac{\omega_{l}}{2}+i\mu_2)+[-i\frac{\bar{j}_{2}\pi}{N\tau_0}+i\frac{\omega_{l}}{2}+i\mu_1+\frac{\Gamma}{2}(\frac{2N}{1-e^{i2\bar{j}_{2}\pi/N}}-N)]}
\label{2CeAnaNinf},
\end{align}\end{widetext}
where $\bar j_1$ and $\bar j_2$ are two solutions satisfying the transcendental equation in Eq.~(\ref{omegak1inf}) [see Fig.~\ref{y1y2}(c)]. The long-time dynamics in Eqs.~(\ref{CxAnaNinf})-(\ref{2CeAnaNinf}) is consistent with that given in Eqs.~(\ref{CxAnaN}), (\ref{CeAnaN}), (\ref{2CxAnaN}), and (\ref{2CxAnaN}). In Fig.~\ref{inStaticBS}, we show how the static bound state is formed. We plot the time evolution of the field intensity function $\bar{p}(x,t)$ [Fig.~\ref{in3D3Dcross}(a) and (c)] and the long-time field intensity distribution $\bar{P}(x,t\rightarrow \infty)$ [Fig.~\ref{in3D3Dcross}(b) and (d)] of a giant atom with $N = 3$ coupling points.
\begin{figure}
\includegraphics[width=8.2cm]{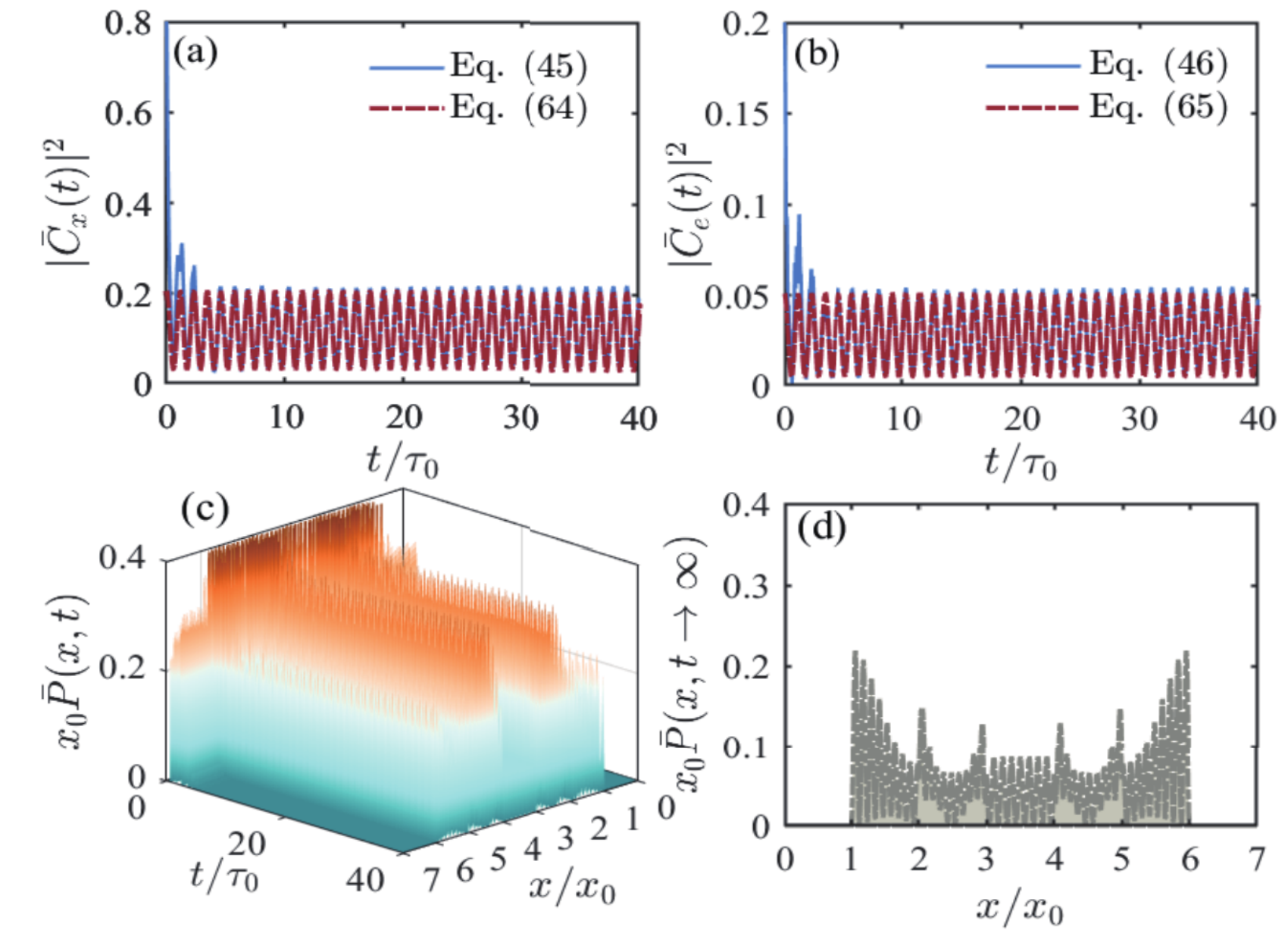}
\vspace{-0.01cm}
\caption{\label{incase2} (Color online) Periodic equal-amplitude oscillating bound states for the three-level giant atom with the number of coupling points $N = 6$. The parameters $\bar{j_1}=23$ and $\bar{j_2}=26$ are chosen from Fig.~\ref{y1y2}(c). In Fig.~\ref{incase2}(a) and (b), the red-dashed line shows the analytical solution by Eqs.~(\ref{2CxAnaNinf}) and (\ref{2CeAnaNinf}), while the blue-solid line denotes the numerical solution by Eqs.~(\ref{dotcxinf}) and (\ref{dotceinf}). Figure~\ref{incase2}(c) corresponds to the field intensity $\bar{P}(x,t) = |\bar{\phi}(x,t)|^2$ in Eq.~(\ref{phiinf}) as a function of time $t$ and position $x$. Figure~\ref{incase2}(d) represents the field intensity distribution $\bar{P}(x,t)$ in the photonic waveguide at $t\rightarrow \infty$. We take $G \tau_0= 2.8\pi$, $\omega_x \tau_0=7.0366\pi$, $\omega_e \tau_0=0.2 \pi$, $\omega_l \tau_0=2.5 \pi$, $\Gamma \tau_0=0.1825\pi$, ${\bar{C}_x}(0) = \sqrt {0.8} $, and ${\bar{C}_e}(0) = \sqrt {0.2}$.}
\end{figure}

Similar to the case of the static bound state, we show the dependence of the field intensity function $\bar{p}(x,t)$ on position $x$ and time $t$ in Fig.~\ref{incase2} for the periodic equal-amplitude oscillating bound states. From Figs.~\ref{inStaticBS}, \ref{in3D3Dcross} and \ref{incase2}, it can be seen that for the same parameters, the analytical results of the driven three-level giant atom coupling with an infinite photonic waveguide are consistent with those of the driven three-level giant atom coupling with a semi-infinite photonic waveguide for the long-time dynamics. The numerical results for the driven three-level giant atom coupling with a semi-infinite photonic waveguide are shown in Fig.~\ref{inStaticBS} to oscillate apparently a little more regularly, which is due to the fact that the photons have not been reflected in the photonic waveguide.

\section{NON-MARKOVIAN DYNAMICS FOR N DRIVEN NONINTERACTING THREE-LEVEL GIANT ATOMS SYSTEM}\label{Sec8}
\begin{figure*}[t]
\centering\scalebox{0.35}{\includegraphics{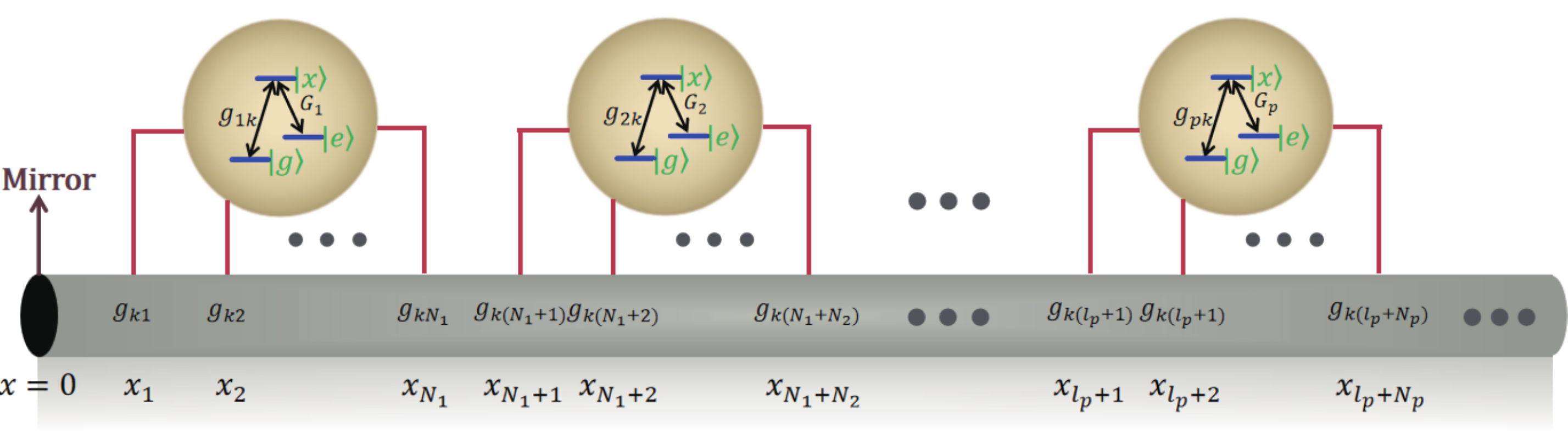}}
\vspace{-0.5cm}
\caption{One-dimensional semi-infinite photonic waveguide coupling with an array of noninteracting driven (with the frequency $\omega_l$ and driving strength $G_p$) three-level giant atoms (with levels $|x\rangle$, $|e\rangle$, and $|g\rangle$ and eigenfrequencies $\omega_x$, $\omega_e$, and $\omega_g$) via $ \sum_{p = 1}^P {{N_p}} $ coupling points with coupling strength ${g_{kn}} = \sqrt {\Gamma \upsilon/\pi }\sin (kn{x_0})$ and position $x_n$, where $N_p$ is the number of coupling points for the $q$th giant atom. $l_p$ represents the sum of the number of all coupling points preceding the first coupling point of the $p$th giant atom, which is expressed as $l_p = \sum_{p'=1}^{p - 1} {{N_{p'}}} $. The separation between any two adjacent coupling points is $x_0$, which is equal to that between the first coupling point and the mirror.}\label{model3}
\end{figure*}
In this section, we extend $P$ driven noninteracting three-level giant atoms coupled to a one-dimensional semi-infinite photonic waveguide. The model under study comprises a one-dimensional semi-infinite photonic waveguide lying along the $x$ axis, including $P$ non-interacting atoms placed at $x_p = p{x_0}$ ($p = 1,2, \cdots P$). $l_p$ represents the sum of the number of all coupling points preceding the first coupling point of the $p$th giant atom, which is denoted as $l_p = \sum_{p'=1}^{p - 1} {{N_{p'}}} $. The total Hamiltonian of the system in the RWA can be written as
\begin{equation}
\begin{aligned}
\hat H = {{\hat H}_F} + {{\hat H}_{AF}} + {{\hat H}_{AW}},
\label{noninteractingHT}
\end{aligned}
\end{equation}
with
\begin{equation}
\begin{aligned}
{{\hat H}_F} =& \sum\limits_{p = 1}^P{({\omega _{p,x}}}  - \frac{{{\omega _{l}}}}{2}){{\hat \sigma }_{p,xx}}+ \sum\limits_{p = 1}^P {({\omega _{p,e}}}  + \frac{{{\omega _{l}}}}{2}){{\hat \sigma }_{p,ee}}\\
& +\int_0^{{k_c}} {dk{{(\Omega }_k-\omega_l/2)}\hat a_k^\dag } {{\hat a}_k},\\
{{\hat H}_{AF}} =& \sum\limits_{p = 1}^P {{G _p}} ({{\hat \sigma }_{p,ex}} + {{\hat \sigma }_{p,xe}}),\\
{{\hat H}_{AW}} =&\sum_{p=1}^P\sum_{n=l_p+1}^{l_p+N_p}\int_0^{k_c}dk(g_{kn}\hat{a}_k^\dagger\hat\sigma_{p,gx}+g_{kn}^*\hat{a}_k\hat\sigma_{p,xg}),
\label{Hgeneral}
\end{aligned}
\end{equation}
where ${\hat\sigma _{p,df}} = \left| d \right\rangle_{p}\left\langle f \right|$ ($d,f = x,g,e$) is Pauli operator of the $p$th atom. The first equation for ${\hat H}_F$ of Eq.~(\ref{Hgeneral}) is the free Hamiltonian of $N$ three-level giant atoms and the semi-infinite photonic waveguide, respectively. The level frequencies of the $p$th giant atom are represented by ${\omega _{p,x}}$ and ${\omega _{p,e}}$. The second equation for ${\hat H}_{AF}$ of Eq.~(\ref{Hgeneral}) denotes the interaction between the $p$th giant atom and the $p$th external classical field, where ${G_p}$ denotes the $p$th driving strength of the $p$th classical field with the same driving frequency $\omega_l$ for all the atoms. The last equation for ${\hat H}_{AW}$ of Eq.~(\ref{Hgeneral}) is the interaction between the $p$th atom and photonic waveguide. The initial state is considered to be ${\left| {\psi_P(0)} \right\rangle } = \sum\nolimits_{p = 1}^P {[{C_{p,x}}(0)\left| {{X_p},0} \right\rangle  + {C_{p,e}}(0)} \left| {{E_p},0} \right\rangle ]$,
where $\left| {{X_p},0} \right\rangle $ and $\left| {{E_p},0} \right\rangle $ represent that the $p$th giant atom respectively is on the excited states $\left| x \right\rangle $ and $\left| e \right\rangle $, but simultaneously  the photon in the vacuum state $|0\rangle$ of the photonic waveguide field, i.e., $\left| {{X_p},0} \right\rangle  = {\left| g \right\rangle _1} \otimes {\left| g \right\rangle _2} \otimes  \cdots  \otimes {\left| x \right\rangle _p} \otimes  \cdots  \otimes {\left| g \right\rangle _N} \otimes {\left| 0 \right\rangle _1} \otimes {\left| 0 \right\rangle _2} \otimes  \cdots  \otimes {\left| 0 \right\rangle _\infty }$ and $\left| {{E_p},0} \right\rangle  = {\left| g \right\rangle _1} \otimes {\left| g \right\rangle _2} \otimes  \cdots  \otimes {\left| e \right\rangle _p} \otimes  \cdots  \otimes {\left| g \right\rangle _N} \otimes {\left| 0 \right\rangle _1} \otimes {\left| 0 \right\rangle _2} \otimes  \cdots  \otimes {\left| 0 \right\rangle _\infty }$. The total Hamiltonian conserves the number of excitations so that the state of the system at time $t$ reads
\begin{eqnarray}
\left| {\psi (t)} \right\rangle  &=& \sum\limits_{p = 1}^P {{C_{p,x}}(t)\left| {{X_p},0} \right\rangle }  + \sum\limits_{p = 1}^P {{C_{p,e}}(t)\left| {{E_p},0} \right\rangle } \nonumber\\
&& + {\int {dk{\beta _{k}}(t)\left| {{\tilde g},{1_k}} \right\rangle } },
\label{psigeneral}
\end{eqnarray}
where $\left| {\tilde g,{1_k}} \right\rangle  = {\left| g \right\rangle _1} \otimes {\left| g \right\rangle _2} \otimes  \cdots  \otimes {\left| g \right\rangle _N} \otimes {\left| 0 \right\rangle _1} \otimes {\left| 0 \right\rangle _2} \otimes  \cdots  \otimes {\left| 1 \right\rangle _k} \otimes  \cdots  \otimes {\left| 0 \right\rangle _\infty }$. The first two terms of Eq.~(\ref{psigeneral}) mean that except for the $p$th giant atom that is in the excited state $\left| x \right\rangle $ or $\left| e \right\rangle $ with the probabilities ${\left| {{C_{p,x}}(t)} \right|^2}$ and ${\left| {{C_{p,e}}(t)} \right|^2}$, all the other giant atoms are in the ground state $\left| g \right\rangle $. The last term of Eq.~(\ref{psigeneral}) represents the environment only one excitation in mode $k$ with the probability ${\left| {{\beta _k}(t)} \right|^2}$ with $\beta_{k}(0)=0$, but all other modes in the environment are in a vacuum. Bringing Eq.~(\ref{noninteractingHT}) and (\ref{psigeneral}) into Schr${\rm{\ddot o}}$dinger equation $\frac{\partial}{\partial t}\left| {{\psi (t)}} \right\rangle=-i\hat H\left| {{\psi (t)}} \right\rangle$, which transforms $\left| {{\psi (t)}} \right\rangle$ into three coupled equations of motion for the three-level giant atom and field excitation amplitudes as follows
\begin{eqnarray}
{{\dot C}_{p,x}}(t) &=&  - i({\omega _{p,x}} - \frac{{{\omega _l}}}{2}){C_{p,x}}(t) - i{G _p}{C_{p,e}}(t) \nonumber \\
&&-i\sum\limits_{n = l_p + 1}^{l_p + {N_p}} \int {dk{g^{*}_{kn}}} {\beta_k}(t),\label{differentcx}\\
{{\dot C}_{p,e}}(t) &=&  - i({\omega _{p,e}}+\frac{{{\omega _l}}}{2}){C_{p,e}}(t) - i{G_p}{C_{p,x}}(t) ,\label{differentce}\\
{{\dot \beta }_k}(t) &=&  - i(\Omega _k-\frac{{{\omega _l}}}{2})\beta _k(t) - i\sum\limits_{p = 1}^N \sum\limits_{n = l_p + 1}^{l_p + {N_p}} {{g_{kn}}{C_{p,x}}(t)}.\nonumber \\ \label{differentbk}
\end{eqnarray}
By integrating Eq.~(\ref{differentbk}), we obtain
\begin{align}
\beta(k,t)=&-i\sqrt{\frac{\Gamma v}{\pi}}\int_0^t\sum_{p=1}^P\sum_{n=l_{p+1}}^{{l_p}+{N_p}}\nonumber\\
&\cdot\sin(knx_0)C_{p,x}(s)e^{-i(\Omega _k-\frac{\omega_l}{2})(t-s)}ds,
\label{betakn}
\end{align}
where the coupling coefficient between the $p$th giant atom and photonic waveguide can be given by ${g_{kn}} = \sqrt {\Gamma \upsilon/\pi }\sin (kn{x_0})$ with photon group velocity $\upsilon $ and decay rate $\Gamma $. Substituting Eq.~(\ref{betakn}) into Eq.~(\ref{differentcx}), we get
\begin{align}
{{\dot C}_{p,x}}(t) =&- i({\omega _{p,x}} - \frac{{{\omega _l}}}{2}){C_{p,x}}(t) - i{G _p}{C_{p,e}}(t) \nonumber\\
 &- \frac{\Gamma }{2}\sum\limits_{\tilde p = 1}^P {\sum\limits_{m,n = {l_{\tilde p}} + 1}^{{l_{\tilde p}} + {N_{\tilde p}}} {{C _{\tilde p,x}}(t - \left| {m - n} \right|{\tau _0})} } \nonumber \\
 &\cdot \Theta (t - \left| {m - n} \right|{\tau _0}) \nonumber\\
 &+ \frac{\Gamma }{2}\sum\limits_{\tilde p = 1}^P {\sum\limits_{m,n = {l_{\tilde p}} + 1}^{{l_{\tilde p}} + {N_{\tilde p}}} {{C _{\tilde p,x}}\left[ {t - \left( {m + n} \right){\tau _0}} \right]} } \nonumber \\
 &\cdot \Theta \left[ {t - \left( {m + n} \right){\tau _0}} \right],\label{cqxdaoshu}\\
 {{\dot C}_{p,e}}(t) =&- i({\omega _{p,e}} + \frac{{{\omega _l}}}{2}){C_{p,e}}(t) - i{G _p}{C_{p,x}}(t).\label{cqedaoshu}
\end{align}
In particular, for $N$ driven three-level giant atoms, the set of differential equations for their coupling with an infinite photonic waveguide becomes
\begin{align}
{{\dot C}_{p,x}}(t) =&  - i({\omega _{p,x}} - \frac{{{\omega _l}}}{2}){C_{p,x}}(t) - i{G _p}{C_{p,e}}(t) \nonumber\\
 &- \frac{\Gamma }{2}\sum\limits_{\tilde p = 1}^P {\sum\limits_{m,n = {l_{\tilde p}} + 1}^{{l_{\tilde p}} + {N_{\tilde p}}} {{C _{\tilde p,x}}(t - \left| {m - n} \right|{\tau _0})} } \nonumber \\
 &\cdot \Theta (t - \left| {m - n} \right|{\tau _0}), \nonumber\\
\label{cqxdaoshu}\\
 {{\dot C}_{p,e}}(t) =&- i({\omega _{p,e}} + \frac{{{\omega _l}}}{2}){C_{p,e}}(t) - i{G _p}{C_{p,x}}(t).\label{cqedaoshu}
\end{align}
By solving the set of time-delay differential equations given by Eqs.~(\ref{cqxdaoshu}) and (\ref{cqedaoshu}), we can get the complete information of the $P$ driven noninteracting three-level giant atoms. The interaction between any two or more giant atoms induced by the semi-infinite photonic waveguide generally exists, which may result in the quantum correlation between different driven three-level giant atoms. We find that the results given by Eqs.~(\ref{cqxdaoshu}) and (\ref{cqedaoshu}) are reduced to the previous ones given by Eqs.~(\ref{dotcx}) and (\ref{dotce}) for the case of $N = 1$. Different from the case of single giant atom, we show that there will be more bound states (exceeding two bound states) in the photonic waveguide simultaneously, which provides an active way for us to control the coupled giant atoms via engineering the non-Markovian environment. We will not discuss the problem in detail here, in which readers who are interested in Eqs.~(\ref{cqxdaoshu}) and (\ref{cqedaoshu}) can try it out.

\section{conclusions and discussions}\label{Sec9}
In this paper, we have studied the non-Markovian dynamics in the spontaneous emission of a three-level giant atom interacting with a one-dimensional semi-infinite photonic waveguide and driven by a classical driving field. By Laplace transform, we obtained the analytical solutions for the atomic probability amplitudes, which reflect nonexponential dissipations due to photon transfer between multiple coupling points and reabsorption after reflection in the semi-infinite photonic waveguide. We discussed the origins and two independent conditions for the formation of bound states and derived two different types of bound states. According to the number of bound states in the photonic waveguide, the bound states are divided into the static case with one bound state and periodic equal-amplitude oscillation with two bound states. The periodic equal-amplitude oscillating bound states display the period oscillation behavior for the giant atom dynamics. Oscillatory bound states with multiple bound states provide a method for us to store and manipulate more complex quantum information.  Because of no mirror at one end of the infinite photonic waveguide, there is one purely imaginary solution for the complex frequency leading to this case compared to that of a driven three-level giant atom coupling with a semi-infinite photonic waveguide, which implies that there is only one bound state condition. Under this condition, two different kinds of bound states also can be found (the static bound state and the periodic equal-amplitude oscillating bound state).

The studies of non-Markovian dynamics in three-level giant atoms coupling with a one-dimensional semi-infinite photonic waveguide may open up a method to better understand non-Markovian quantum networks and quantum communications. As a prospect, it will be interesting to investigate the driven three-level giant atoms coupling with an infinite or semi-infinite photonic waveguide, where the distances between neighbouring coupling points are unequal. The Hamiltonian of this system is described by
\begin{equation}\begin{aligned}
\hat{H}_{uneq}=&\mu_1 \hat\sigma_{xx}+\mu_2\hat\sigma_{ee}
\\
&+\int_{-\infty}^{+\infty}(\Omega_{k}-\frac{\omega_{l}}{2})\hat{a}_k^\dagger\hat{a}_kdk+G(\hat\sigma_{ex}+\hat\sigma_{xe}) \\ &+\sum_{n=1}^{N}\int_{-\infty}^{+\infty}dk(g_{kn}\hat{a}_k^\dagger\hat\sigma_{gx}+g^*_{kn}\hat{a}_k\hat\sigma_{xg}),
\label{hatH2}\end{aligned}\end{equation}
where $g_{kn}=\sqrt{\Gamma\upsilon/\pi}\sin(kx_n)$ or $g_{kn}=\sqrt{\Gamma\upsilon/2\pi}e^{ikx_n}$. There are two cases that the interval of the coupling point is an arithmetic progression $x_{n}=nx_0+n(n-1)d/2$, or the interval of the coupling point is a geometric progression $x_{n}=x_0(1-q^{n})/(1-q)$. Other forms of $x_n$ that are regular or disordered with respect to $n$ can also be considered. Moreover, we also can further study the anisotropic nonrotating wave driven three-level giant atom systems via applying the methods in Refs.\cite{Shi1536022018,Lo0638072018,Shen0237072022,Shen0238562018,Xie0210462014,Chen0437082021,Rodriguez0466082008,Nakajima43631955,Frohlich8451950,Frohlich2911952,Lu0543022007}. The total Hamiltonian of this system in this case is
\begin{equation}\begin{aligned}
\hat{H}_{aniso}=& \begin{aligned}\omega_x|x\rangle\langle x|+\omega_e|e\rangle\langle e|+\int_0^{k_c}dk\Omega_k\hat{a}_k^\dagger\hat{a}_k\end{aligned}
\\
&+Ge^{i\omega_lt}\hat\sigma_{ex}+Ge^{-i\omega_lt}\hat\sigma_{xe} \\
&+\sum_{m=1}^{M}\int_0^{k_c}dk [A_{km}(\hat{a}_k^\dagger\hat\sigma_{gx}+\hat{a}_k\hat\sigma_{xg})\\
&+ {B_{km}}( {\hat a_k^\dag {\hat\sigma _{xg} } + {{\hat a}_k}{\hat\sigma _{gx} }} )] ,
\end{aligned}\end{equation}
where $A_{km}$ and $B_{km}$ represent the coupling strengths of the rotating-wave and non-rotating-wave interactions, respectively.

\section*{ACKNOWLEDGMENTS}
This work was supported by National Natural Science Foundation of China under Grants No. 12274064 and Scientific Research Project for Department of Education of Jilin Province under Grant No. JJKH20241410KJ.

\appendix
\section{DERIVATION OF EQ.~(\ref{dotcx})}\label{A}
Calculating the differential equation in Eq.~(\ref{dotbetak}), we can get
\begin{align}
\beta_{k}(t)=&e^{-i(\Omega_{k}-\frac{\omega_{l}}{2})(t-0)}\beta_{k}(0) \nonumber \\
&-i\int_{0}^{t}C_{x}(s)\sum_{n=1}^{N} g_{kn}e^{-i(\Omega_{k}-\frac{\omega_{l}}{2})(t-s)}ds.
\label{A1}\end{align}
By substituting the coupling strength $g_{kn}=\sqrt{\Gamma\upsilon/\pi}\sin(knx_0)$ and  initial condition $\beta_{k}(0)=0$ into Eq.~(\ref{A1}), we can get Eq.~(\ref{betak}). Substituting Eq.~(\ref{betak}) into Eq.~(\ref{dotCx}) obtains
\begin{widetext}\begin{equation}
\dot{C_{x}}\left(t\right)=-i\mu_1C_{x}(t)-iGC_{e}\left(t\right)-\frac{\Gamma v}{\pi}\sum_{m,n=1}^{N}\int dk\sin(kmx_0)\sin(knx_0)\int_0^{t}C_{x}\left(s\right)e^{-i(\Omega_{k}-\frac{\omega_{l}}{2})\left(t-s\right)}ds
.\label{A2}\end{equation}
With
\begin{equation}
\sin(kmx_0)\sin(knx_0)=\frac{1}{2}\left\lbrace\frac{1}{2}\left\lbrack e^{ik(m-n)x_0}+e^{-ik(m-n)x_0}\right\rbrack-\frac{1}{2}\left\lbrack e^{ik(m+n)x_0}+e^{-ik(m+n)x_0}\right\rbrack\right\rbrace
,\label{sin}\end{equation}
and the identity  $\frac{1}{2\pi}\int_{-\infty}^{\infty}e^{ik(x-a)}dk=\delta(x-a)$, Eq.~(\ref{A2}) is reduced to
\begin{align}
\dot{C}_{x}(t)  =&-i\mu_1 C_{x}(t)-iG C_{e}(t)\nonumber \\   &-\frac{\Gamma}{2}\sum_{m,n=1}^{N}\int_0^{t}C_{x}(s)e^{i\frac{\omega_{l}}{2}(t-s)}\delta[s-(t-(m-n)\tau_0)]ds\nonumber \\  &-\frac{\Gamma}{2}\sum_{m,n=1}^{N}\int_0^{t}C_{x}(s)e^{i\frac{\omega_{l}}{2}(t-s)}\delta[s-(t+(m-n)\tau_0)]ds\nonumber \\  &+\frac{\Gamma}{2}\sum_{m,n=1}^{N}\int_0^{t}C_{x}(s)e^{i\frac{\omega_{l}}{2}(t-s)}\delta[s-(t-(m+n)\tau_0)]ds\nonumber \\  &+\frac{\Gamma}{2}\sum_{m,n=1}^{N}\int_0^{t}C_{x}(s)e^{i\frac{\omega_{l}}{2}(t-s)}\delta[s-(t+(m+n)\tau_0)]ds
.\label{sum}\end{align}\end{widetext}
Eq.~(\ref{dotcx}) can be obtained by utilizing the Heaviside step function and $\int f(x){\delta(x-a)}dx=f(a)$ plus $\int_{0}^{a} f(x){\delta(x-a)}dx={\frac{1}{2}}f(a)$ \cite{Gardiner2000} to Eq.~(\ref{sum}).
\\
\\
\section{DERIVATION OF EQS.~(\ref{im}) AND ~(\ref{re})}\label{B}
\begin{widetext}
In this Appendix, we derive Eqs.~(\ref{im}) and ~(\ref{re}). In our scheme, the coupling points are equidistant with time delay $\tau _0$ between neighboring coupling points. Therefore, all the possible time delays can be written in the form $\left| {m - n} \right|{\tau _0} = n{\tau _0}$ with ${n = 0, 1,\cdots, N - 1}$. The combination number is $N$ for $n = 0$ and $2\left( {N - n} \right)$ for $n \ne 0$. Therefore, we can expand the sum terms of Eq.~(\ref{sk}) as
\begin{align}
\sum_{m,n=1}^{N}e^{-\left(s_{j}-\frac{i\omega_{l}}{2}\right)|m-n|\tau_0}&=2\frac{N-(N+1)e^{-\left(s_{j}-\frac{i\omega_{l}}{2}\right)\tau_0}+e^{-(N+1)\left(s_{j}-\frac{i\omega_{l}}{2}\right)\tau_0}}{[1-e^{-\left(s_{j}-\frac{i\omega_{l}}{2}\right)\tau_0}]^2}-N\equiv A,\label{suma}
\\ \sum_{m,n=1}^{N}e^{-\left(s_{j}-\frac{i\omega_{l}}{2}\right)\left(m+n\right)\tau_0}&=\frac{e^{-2\left(s_{j}-\frac{i\omega_{l}}{2}\right)\tau_0}[1-e^{-\left(s_{j}-\frac{i\omega_{l}}{2}\right)N\tau_0}]^2}{[1-e^{-\left(s_{j}-\frac{i\omega_{l}}{2}\right)\tau_0}]^2}\equiv B.\label{sumb}
\end{align}
Thus, with Eq.~(\ref{sk}) by setting $s_{\alpha}\rightarrow s_j=-i\omega_j$, we have
\begin{eqnarray}
[-i\omega_{j}+i\mu_1+\frac\Gamma2(A-B)](-i\omega_{j}+i\mu_2)
+G^2=0.\label{A3}
\end{eqnarray}
Two expressions in Eqs.~(\ref{im}) and (\ref{re}) correspond to the imaginary and real parts of Eq.~(\ref{A3}), respectively.
\end{widetext}

\section{DERIVATION OF EQS.~(\ref{CxAnaN}),~(\ref{CeAnaN}),~(\ref{CxAnaN1}), AND ~(\ref{CeAnaN1})}\label{C}

\begin{widetext}
Because of $\tilde s_{j}=s_j-i\omega_l/2=-i2j\pi/N\tau_0$, $e^{-\tilde s_{j}N\tau_0}=1$, $e^{\tilde s_{j}N\tau_0}=1$, with Eqs.~(\ref{suma}) and (\ref{sumb}), Eq.~(\ref{kai1se}) can be simplified into
\begin{align}\begin{aligned}
\chi_{1}(\tilde s_j)&=\frac{\Gamma}{2}\sum_{m,n=1}^{N}\left[e^{-\tilde s_{j}|m-n|\tau_0}-e^{-\tilde s_{j}(m+n)\tau_0}\right]=\frac{\Gamma}{2}\left(\frac{2N}{1-e^{i2j\pi/N}}-N\right),
\label{B4}\end{aligned}\end{align}
and Eq.~(\ref{kai2se}) becomes
\begin{align}
\chi_{2}(\tilde s_j)=\frac{\Gamma}{2}\sum_{m,n=1}^{N}\left[(m+n)\tau_0e^{-\tilde s_{j}(m+n)\tau_0}-|m-n|\tau_0e^{-\tilde s_{j}|m-n|\tau_0}\right]=\frac{d}{d\tilde s_{j}}\chi_{1}(\tilde s_j),
\end{align}
which leads to
\begin{align}
\chi_{2}(\tilde s_j)=\frac{\Gamma}{2}\left[\frac{4N\tau_0e^{-2\tilde s_{j}\tau_0}-4N\tau_0e^{-\tilde s_{j}\tau_0}}{(1-e^{-\tilde s_{j}\tau_0})^3}\right]=\frac{\Gamma}{2}\frac{N\tau_0}{\sin^2(\frac{j\pi}{N})}.
\label{B6}\end{align}
For the case $\tilde s_{j}=-i2j\pi/N\tau_0$, bringing Eqs.~(\ref{B4}) and (\ref{B6}) into Eqs.~(\ref{Cxtsk}) and (\ref{Cetsk}) gives Eqs.~(\ref{CxAnaN}) and (\ref{CeAnaN}). Calculating the case $\tilde s_{j}=-i2k\pi/(N+1)\tau_0$ in the same way, we can get
\begin{align}
\chi_{1}(\tilde s_j)&=\frac{\Gamma}{2}\left[\frac{2(N+1)}{1-e^{i2j\pi/(N+1)}}-(N+1)\right], \label{B7} \\
\chi_{2}(\tilde s_j)&=\frac{\Gamma}{2}\frac{(N+1)\tau_0}{\sin^2(\frac{j\pi}{N+1})}.\label{B8}
\end{align}
Substituting Eqs.~(\ref{B7}) and (\ref{B8}) into Eqs.~(\ref{Cxtsk}) and (\ref{Cetsk}) leads to Eqs.~(\ref{CxAnaN1}) and (\ref{CeAnaN1}).
\end{widetext}

\section{DERIVATION OF EQS.~(\ref{dotcxinf}) and (\ref{dotceinf})}\label{D}
\begin{widetext}
In this Appendix, we derive a set of delay differential equations based on Eq.~(\ref{hatHinf}) in the infinite-photonic waveguide case
\begin{align}
\dot{\bar{C}}_{x}(t)=&-i\mu_1\bar{C}_{x}(t)-iG\bar{C}_{e}(t)-i\sum_{n=1}^{N}\int_{-\infty}^{+\infty}dk\bar g_{kn}\bar{\beta}_{k}(t),
\label{d1}\\
\dot{\bar{C}}_{e}(t)=&-i\mu_2 \bar{C}_{e}(t)-iG\bar{C}_{x}(t),
\label{d2}\\
\dot{\bar{\beta}}_{k}(t)=&-i(\Omega_{k}-\frac{\omega_{l}}{2})\bar{\beta}_{k}(t)-i\bar{C}_{x}(t)\sum_{n=1}^{N} \bar g_{kn}.
\label{d3}
\end{align}
Integrating Eq.~(\ref{d3}) with $\bar{g}_{kn}=\sqrt{\Gamma v/2\pi}e^{iknx_0}$, we get
\begin{align}
\bar{\beta}_{k}(t)=&-i\sqrt{\frac{\Gamma\upsilon}{2\pi}}\int_{0}^{t}\bar{C}_{x}(s)\sum_{n=1}^{N}e^{iknx_{0}}e^{-i(\Omega_{k}-\frac{\omega_{l}}{2})(t-s)}ds
.\label{d4}
\end{align}
Substituting Eq.~(\ref{d4}) into Eq.~(\ref{d1}) gives
\begin{align}
\dot{\bar{C}}_{x}\left(t\right)=&-i\mu_1\bar{C}_{x}(t)-iG\bar{C}_{e}\left(t\right)\nonumber\\
&-\frac{\Gamma v}{4\pi}\sum_{m,n=1}^{N}\int dke^{ik(m-n)x_{0}}\int_0^{t}\bar{C}_{x}\left(s\right)e^{-i(\Omega_{k}-\frac{\omega_{l}}{2})\left(t-s\right)}ds\nonumber\\
&-\frac{\Gamma v}{4\pi}\sum_{m,n=1}^{N}\int dke^{ik(n-m)x_{0}}\int_0^{t}\bar{C}_{x}\left(s\right)e^{-i(\Omega_{k}-\frac{\omega_{l}}{2})\left(t-s\right)}ds
.\label{d5}\end{align}
Using the identity  $\frac{1}{2\pi}\int_{-\infty}^{\infty}e^{ik(x-a)}dk=\delta(x-a)$, we can get
\begin{align}
\dot{\bar{C}}_{x}(t)=&-i\mu_1 \bar{C}_{x}(t)-iG \bar{C}_{e}(t)\nonumber \\   &-\frac{\Gamma}{2}\sum_{m,n=1}^{N}\int_0^{t}\bar{C}_{x}(s)e^{i\frac{\omega_{l}}{2}(t-s)}\delta[s-(t-(m-n)\tau_0)]ds\nonumber \\
&-\frac{\Gamma}{2}\sum_{m,n=1}^{N}\int_0^{t}\bar{C}_{x}(s)e^{i\frac{\omega_{l}}{2}(t-s)}\delta[s-(t+(m-n)\tau_0)]ds
.\label{d6}\end{align}
Equation~(\ref{dotcxinf}) can be obtained by applying the Heaviside step function and $\int f(x)\delta(x-a)dx=f(a)$ plus $\int_{0}^{a} f(x){\delta(x-a)}dx={\frac{1}{2}}f(a)$ \cite{Gardiner2000} to Eq.~(\ref{d6}).
\end{widetext}

\end{document}